\documentclass[]{emulateapj}

\usepackage{wasysym} % for astrosun
\usepackage{amssymb}
\usepackage{lscape}
\usepackage{color}

\begin{document}

\title{X-Ray Insights Into the Physics of Mini-BAL Quasar Outflows\footnote{Accepted to ApJ. (c) Copyright 2009.  The American Astronomical Society.  All rights reserved.  Printed in U.S.A.}}

\author{Robert R. Gibson\altaffilmark{1}, W.~N. Brandt\altaffilmark{1}, S.~C. Gallagher\altaffilmark{2}, and Donald P. Schneider\altaffilmark{1}}
\email{rgibson@astro.washington.edu}

\altaffiltext{1}{Department of Astronomy and Astrophysics, Pennsylvania State University, 525 Davey Laboratory, University Park, PA 16802}
\altaffiltext{2}{Department of Physics and Astronomy, The University of Western Ontario, 1151 Richmond Street, London, ON N6A 3K7, Canada}

\shorttitle{X-Raying Mini-BAL Quasars}
\shortauthors{Gibson et al.}

% Turn off ``underfull hbox'' warning
\hbadness=10000

\clearpage

\begin{abstract}
We examine the UV and \mbox{X-ray} properties of 256 radio-quiet SDSS quasars (QSOs) observed in \mbox{X-rays} with {\it Chandra} and/or {\it XMM-Newton} in order to study the relationship between QSOs with broad \ion{C}{4} absorption lines (BALs; width $> 2000$~km~s$^{-1}$) and those with \ion{C}{4} mini-BALs (here defined to have widths of 1000--2000~km~s$^{-1}$).  Our sample includes 42 BAL and 48 mini-BAL QSOs.  The relative \mbox{X-ray} brightness and hard spectral slopes of the mini-BAL population are, on average, intermediate between those of BAL and non-BAL QSOs, as might be expected if narrower and broader absorption line outflows are physically related.  However, a significant population of mini-BALs has outflow velocities higher than would be expected for BAL QSOs of the same relative \mbox{X-ray} brightness.  Consistenly strong \mbox{X-ray} absorption is apparently not required to accelerate at least some mini-BALs to high outflow velocities.  Assuming the mini-BAL features are correctly attributed to intrinsic \ion{C}{4} absorption, we suggest that their observed properties may be explained if mini-BALs are ``seeds'' which can be accelerated to form BALs when sufficient \mbox{X-ray} shielding is present.

We also examine several QSOs with broad \ion{C}{4} absorption that have been recently reported to be unusually \mbox{X-ray} bright.  Such cases are frequently mini-BAL QSOs, which as a population are generally brighter in \mbox{X-rays} than BAL QSOs.  Pointed {\it XMM-Newton} observations also suggest that these sources (or unresolved neighbors) may have been previously observed in a high flux state.
\end{abstract}


\section{INTRODUCTION\label{introSec}}

Broad absorption lines (BALs) are observed in the ultraviolet (UV) spectra of $\approx$15\% of quasars (QSOs) from lines such as \ion{Si}{4}~$\lambda$1400, \ion{C}{4}~$\lambda$1549, \ion{Al}{3}~$\lambda$1857, and \ion{Mg}{2}~$\lambda$2799 \citep[e.g.,][and references therein]{hf03, thrrsvkafbkn06, g+08c}.  In the traditional model, BAL outflows are equatorial disk winds that are ubiquitous in QSOs, and significant \mbox{X-ray} absorption is required to shield the outflow from overionization that would prohibit radiative acceleration \citep[e.g.,][]{mcgv95}.  Indeed, BAL QSOs (i.e., QSOs with BALs evident in their UV spectra) are relatively \mbox{X-ray} weak compared to non-BAL QSOs \citep[e.g.,][]{gsahfbftm95, blw00}, apparently due to strong \mbox{X-ray} absorption \citep[e.g.,][and references therein]{gamwe01, gblemwi01, gbcg02}.  Significant correlations have also been found between the degree of \mbox{X-ray} weakness and acceleration-dependent BAL properties such as outflow velocity and equivalent width, as might be expected for radiatively-accelerated BAL outflows \citep[e.g.,][]{gbcpgs06, g+08c}.

In the current work, we seek to extend analyses of BAL QSO \mbox{X-ray} properties in two respects.  First, we expand the parameter space of BAL studies to include broad absorption features which are not formally considered to be BALs, but may represent an intermediate stage in BAL structure and/or evolution.  The fraction of QSOs with so-called ``mini-BAL'' absorption features (having intermediate widths of 1000--2000~km~s$^{-1}$) is similar to or greater than the fraction with BALs \citep{thrrsvkafbkn06}; this is a large population of sources which has received relatively little attention in \mbox{X-rays}.  Second, we consider targeted {\it XMM-Newton} observations of mini-BAL QSOs in light of our analysis of mini-BAL \mbox{X-ray} properties to investigate previous indications that these sources are unusually \mbox{X-ray} bright.  Such cases could conflict with our current understanding that strong \mbox{X-ray} absorption is needed to radiatively accelerate outflows.

Additionally, this study is one step toward understanding the variety of absorption phenomena that may be classified as ``mini-BALs.''  As our starting point, we identify mini-BALs using a mathematical criterion (\S\ref{optUVDataSec}).  As we discuss further in \S\ref{areMBsIntermSec}, this definition may not distinguish between different types of absorption morphologies that could have a variety of physical origins.  As mini-BAL samples grow, it will be important to determine whether additional formal identification criteria are needed to distinguish new subclasses of absorption phenomena.  Multi-wavelength studies will be valuable to help identify such subclasses.

\subsection{A Continuum of Absorber Properties?\label{contAbsPropSec}}

The velocity-width distinction used to classify absorbers as BALs (width $> 2000$~km~s$^{-1}$), mini-BALs (intermediate widths), and narrow absorption lines (NALs; width $< 500$~km~s$^{-1}$ for \ion{C}{4}) is somewhat arbitrary; it has been suggested that BAL absorption is an extension of the same physical processes responsible for narrower absorption features observed in QSO spectra \citep[e.g., \S4.2.1 of ][]{akdjb99, gb08}.  Previous studies of BAL QSOs have found that their UV absorption properties are correlated with the relative \mbox{X-ray} brightness of these sources \citep[e.g.,][]{gbcpgs06}, indicating a physical link between \mbox{X-ray} absorption and the acceleration of BAL outflows.  QSOs with weaker BALs, which have lower outflow velocities and/or smaller absorption equivalent widths (EWs), are relatively \mbox{X-ray} bright compared to those with stronger BALs \citep{g+08c}.  The next natural step is to determine whether the numerous population of mini-BALs (and NALs) extends the trends observed for BAL QSOs overall.

\mbox{X-ray} studies of smaller samples of mini-BAL QSOs have suggested a link between BAL and mini-BAL absorption.  \citet{gbcg02} measured \mbox{X-ray} absorption in three mini-BAL QSOs similar to the absorption typically observed for BAL QSOs.  The outflow kinetic energy associated with the variable \mbox{X-ray} absorber of the mini-BAL QSO PG~$1115+080$ may be a significant fraction of the bolometric luminosity \citep{cbg03, cbgp07}.  Recently, \citet{mecc08} have studied \mbox{X-ray} observations of three QSOs with intrinsic \ion{C}{4}  NALs and also one source with a \ion{C}{4} absorber velocity width $\la 800$~km~s$^{-1}$, which they classified as a mini-BAL.  The \mbox{X-ray} properties of these sources appeared similar to those of unabsorbed QSOs.  Because of their \mbox{X-ray} brightness and relatively weak UV absorption, they were qualitatively consistent with previous observations that UV absorption strength decreases with QSO \mbox{X-ray} brightness \citep[e.g.,][]{blw00, gbcpgs06, g+08c}.  On the other hand, the NAL/mini-BAL sources of \citet{mecc08} exhibited much higher outflow velocities than would be expected for BAL QSOs having a similar level of \mbox{X-ray} brightness.

The relation between BAL and mini-BAL absorbers has also been addressed by purely optical/UV studies.  \citet{ksgc08} have reported that the distribution of \ion{C}{4} absorption strength is bimodal for the BAL and mini-BAL QSOs cataloged by \citet{thrrsvkafbkn06}, with BAL QSOs preferentially occupying the stronger component of the distribution.  This may be due to physical differences between BAL and mini-BAL QSOs, or it may be a result of viewing a common physical structure at different orientation angles.  Narrower substructures in mini-BALs have been observed to vary in concert, as if the mini-BAL absorber was a single, unified entity \citep{hbj97, meck07}.  However, subcomponents of BALs have been observed to vary independently on multi-year time scales \citep{gbsg08}, raising the possibility that BALs may be partly composed of narrower absorption structures such as mini-BALs.

\subsection{Unusually X-Ray Bright Sources\label{unusXBrightSourcesSec}}

Extremely \mbox{X-ray} bright BAL and mini-BAL QSOs would appear to be incompatible with the requirement that strong \mbox{X-ray} absorption shield the outflow from overionization.  As we show in this work, such exceptional cases are often mini-BAL QSOs, rather than {\it bona fide} BAL QSOs.  Characterizing the general \mbox{X-ray} properties of mini-BAL QSOs will determine whether the \mbox{X-ray} luminosities of these individual sources are truly exceptional for QSOs with absorbers in the mini-BAL regime.  Unusually high \mbox{X-ray} count rates may also have been recorded by previous missions due to source contamination or variability.  Observations at higher angular resolution with {\it Chandra} or {\it XMM-Newton} are therefore essential to secure claims of abnormal \mbox{X-ray} brightness.

The Sloan Digital Sky Survey \citep[SDSS;][]{y+00} Data Release~3 (DR3) BAL catalog \citep{thrrsvkafbkn06} identified 4784 QSOs with broad ($>$1000~km~s$^{-1}$ wide) UV absorption features \citep{thrrsvkafbkn06}.  As part of the {\it XMM-Newton} AO6 observing cycle, we obtained \mbox{X-ray} spectra of two \ion{C}{4} mini-BAL QSOs identified in the catalog that appeared to be anomalously \mbox{X-ray} bright based on their reported {\it ROSAT} count rates.  Recently, \citet{gp08} identified three additional sources with broad absorption features and high {\it ROSAT} count rates.  In the following analysis, we assess the relative brightness of these sources in relation to the overall populations of BAL and mini-BAL QSOs.

\subsection{Conventions\label{conventionsSec}}

Because conventions vary among studies, we briefly describe the terminology used in this work.  BAL QSOs are broadly classified by ionization into low-ionization BAL QSOs (``LoBALs'') and high-ionization BAL QSOs (``HiBALs'').  LoBALs are QSOs that have BALs from ions at lower ionization states such as \ion{Al}{3} or \ion{Mg}{2}.  They may also have BALs from higher ionization stages such as \ion{Si}{4} or \ion{C}{4}.  HiBALs have BALs {\it only} from high ionization stages.  We define velocities flowing outward (with respect to QSO emission rest frames) to be negative.  Positive velocities indicate features that are at {\it longer} wavelengths than the wavelength corresponding to (rest-frame) zero velocity.  However, we use the terms ``greater'' and ``smaller'' velocities to refer to the {\it magnitude} of the velocity, so that an outflow velocity of \mbox{--10,000~km~s$^{-1}$} is ``greater'' than a velocity of \mbox{--5000~km~s$^{-1}$}.

As discussed in \S\ref{optUVDataSec}, we define mini-BALs to have velocity widths of $1000$--$2000$~km~s$^{-1}$; this provides a reliable extension to the BAL regime with widths $>$2000~km~s$^{-1}$.  Narrower features (with widths of \mbox{$500$--$1000$~km~s$^{-1}$}) are classified as mini-BALs in some studies \citep[e.g.,][]{bhs97, mecc08}; however, we do not include such narrow features here to limit contamination from spectral noise and intervening systems.

Unless otherwise noted, wavelengths in this work refer to rest-frame values.  Throughout, we use a cosmology in which $H_0 = 70$~km~s$^{-1}$~Mpc$^{-1}$, $\Omega_M = 0.3,$ and $\Omega_{\Lambda} = 0.7$.

\section{OBSERVATIONS AND DATA REDUCTION\label{obsDataSec}}

\begin{figure} [ht]
  \begin{center}
      \includegraphics[width=5in, angle=-90]{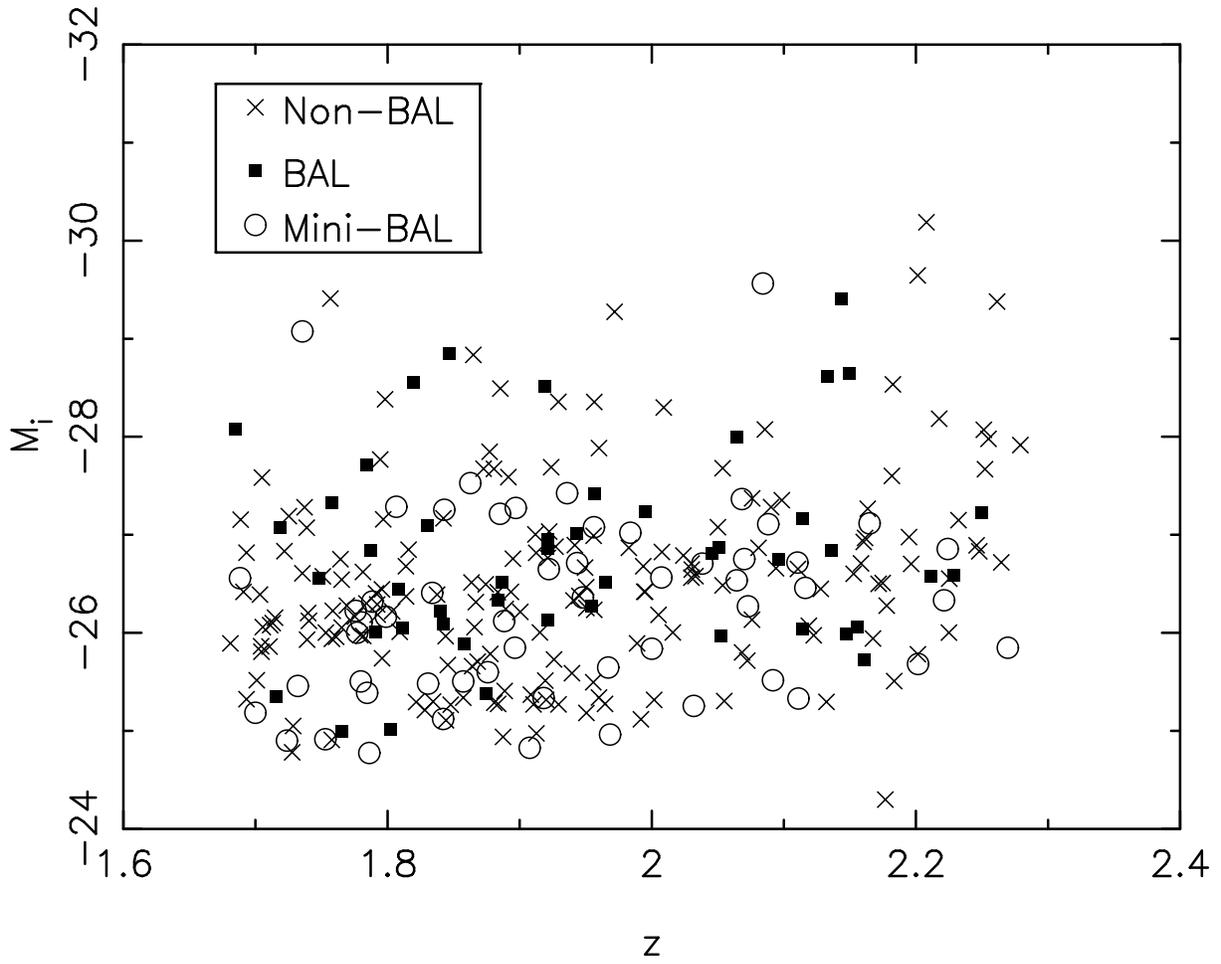}
      \caption{\label{rBTMakeMzPlotFig}Absolute $i$-magnitudes and redshifts reported in the DR5 QSO catalog for the non-BAL (crosses), BAL (filled squares), and mini-BAL (empty circles) QSOs in our sample.}
   \end{center}
\end{figure}

We have obtained SDSS and \mbox{X-ray} spectra for SDSS DR5 QSOs cataloged by \citet{s+07} which also have archived {\it Chandra} or {\it XMM-Newton} \mbox{X-ray} observations publicly available as of 2007 July~26 (for {\it Chandra}) or 2007 August~22 (for {\it XMM-Newton}).  We consider only sources with redshifts $1.68 \le z \le 2.28$ so that we have spectral coverage of the range from approximately 1400 to 2800~\AA.  There are 286 such sources, including both radio-quiet and radio-loud QSOs.  The redshift restriction enables us to test whether our sources show broad absorption lines from low ionization stages such as \ion{Al}{3} and \ion{Mg}{2}.  We discard 2~LoBAL QSOs from the sample, as these sources are known to have \mbox{X-ray} properties that differ from those of HiBALs \citep[e.g.,][]{gbcpgs06}.  We note that the relatively narrow redshift range of our sample, combined with the flux limits for DR5 QSOs, result in a relatively narrow UV luminosity range for our sources.  The absolute $i$-magnitudes, $M_i$, and redshifts, $z$, listed in the DR5 QSO catalog for our sample sources are shown in Figure~\ref{rBTMakeMzPlotFig}.  BAL QSOs, with \ion{C}{4} absorption troughs wider than $2000$~km~s$^{-1}$ (\S\ref{optUVDataSec}), constitute $\approx$16\% of our sources; the fraction of mini-BAL QSOs is $\approx$19\%.  These fractions are not necessarily representative of the full QSO population, as they can be biased by such factors as \mbox{X-ray} telescope observing strategies.

We also obtained observations of \mbox{X-ray} bright mini-BAL QSOs as part of {\it XMM-Newton}~AO6.  These sources were originally identified in the catalog of \citet{thrrsvkafbkn06} to have broad absorption features based on their analysis of spectra available in SDSS Data Release 3.  To test for absorption feature variability, we have also obtained Hobby-Eberly Telescope (HET) Marcario Low Resolution Spectrograph \citep[LRS;][]{hnmtcm98} spectra for these sources.  A log of our targeted observations is given in Table~\ref{obsLogTab}.

Most of the data analysis for this project was performed using the ISIS platform \citep{hd00}.

\subsection{Optical/UV Data Reduction For Target Sources\label{optUVDataSec}}

Before fitting the SDSS DR5 spectra, we multiply them by a single constant to match the photometric $g$, $r$, and $i$ PSF magnitudes \citep{figdss96} to those synthesized from the spectra.  The flux calibration of the HET is less certain, so we do not measure fluxes from our HET spectra.  We convert HET wavelengths to vacuum wavelengths to match the SDSS practice.  We correct all spectra for Galactic extinction using the reddening curve of \citet{ccm89} with the near-UV extension of \citet{o94}.  We obtain $E(B-V)$ from the NASA Extragalactic Database (NED),\footnote{http://nedwww.ipac.caltech.edu/} which uses the dust maps of \citet{sfd98}.

We fit SDSS and HET spectra using the algorithm of \citet{gbsg08}, which we summarize here.  Our continuum model is a power law reddened using the Small Magellanic Cloud reddening curve of \citet{p92}.  We initially fit regions that are generally free from strong absorption or emission features:  1250--1350, 1700--1800, 1950--2200, 2650--2710, and 2950--3700~\AA.  We then iteratively fit the continuum, ignoring at each step wavelength bins that deviate by $>$3$\sigma$ from the current fit in order to exclude strong absorption and emission features.  We fit Voigt profiles to the strongest emission lines expected in the spectrum:  \ion{Si}{4}~$\lambda$1400, \ion{C}{4}~$\lambda$1549, \ion{C}{3}]~$\lambda$1909, \ion{Al}{3}~$\lambda$1857, and \ion{Mg}{2}~$\lambda$2799.  These wavelengths are taken from the SDSS vacuum wavelength list used by the SDSS pipeline to determine emission-line redshifts.\footnote{See http://www.sdss.org/dr6/algorithms/linestable.html and http://www.sdss.org/dr6/algorithms/redshift\_type.html}  We ascribe no physical significance to the Voigt profile; it simply enables us to model emission line cores and wings using a small number of parameters.  We fit emission lines iteratively as well, ignoring at each step bins that are absorbed by more than $2.5\sigma$ from the continuum $+$ emission fit.  Due to the degeneracy between the UV emission continuum shape and the magnitude of intrinsic reddening, we do not attach physical significance to the values of $E(B-V)$ obtained from our fits.  The UV luminosities we report are therefore corrected for Galactic, but not intrinsic, reddening.  The SDSS spectral resolution is $\approx$3~\AA\, while the HET resolution is $\approx$6.1~\AA, based on fits to sky lines.

For sources observed in the FIRST radio survey \citep{bwh95}, we obtain the 1.4~GHz core flux densities from the DR5 QSO catalog.  For sources that were not covered by the FIRST survey, we used 1.4~GHz flux densities obtained from the NVSS survey \citep{ccgyptb98}.  We estimate the monochromatic luminosity at 5~GHz assuming the radio flux follows a power law with a spectral index $\alpha = -0.8$.  We then calculate the radio-loudness parameter \citep[e.g.,][]{sw80, ksssg89},
\begin{eqnarray}
\log(R^*) &\equiv& \log\biggl( \frac{L_{\nu}({\rm 5~GHz})}{L_{2500\mathring{A}}} \biggr).\label{rStarEqn}
\end{eqnarray}
We classify sources with $\log(R^*) \ge 1$ as ``radio loud'' and sources with $\log(R^*) < 1$ as ``radio quiet.''  We classified sources that were not detected in the radio as radio quiet, even in cases where the survey detection threshold (1~mJy for FIRST, $\ga$2.5 mJy for NVSS) did not reach down to $\log(R^*) = 1$.  We estimate that a small number ($\approx 2$) of radio loud QSOs contaminate our samples, but even these are not very radio loud, with upper limits on $\log(R^*) < 1.5$ in almost all cases.  Such sources would not significantly affect our results.

We have calculated the extended balnicity index, $BI_0$, and the absorption index, $AI$, for \ion{C}{4}~$\lambda$1549.  We take the rest wavelength to be that of the red component of the doublet line, 1550.77~\AA\ \citep{vvf96}.  We define $BI_0$ as
\begin{eqnarray}
BI_0 &\equiv& \int^{25,000}_{0} \biggl(1 - \frac{f(-v)}{0.9}\biggr) C~dv,\label{bIDefnEqn}
\end{eqnarray}
where $v$ is the outflow velocity in km~s$^{-1}$ from the rest frame defined by the QSO redshift and $f(v)$ is the ratio of the observed spectrum to the emission model at velocity $v$.  The value $C$ is $0$ unless the observed spectrum has fallen 10\% below the continuum for a velocity width of at least 2000~km~s$^{-1}$ on the red side of the absorption trough, at which point $C$ is set to $1$.  Our definition of $BI_0$ differs from that of $BI$ for \citet{wmfh91} in that we integrate all the way to zero velocity, while the traditional $BI$ measurement integrates only in the range $-25,000$ to $-3000$~km~s$^{-1}$.  This allows us to characterize BAL absorption even at low outflow velocities.

We define the absorption index, $AI$, as
\begin{eqnarray}
AI &\equiv& \int^{29,000}_{0}(1-f(v))C^\prime~dv.\label{aIDefnEqn}
\end{eqnarray}
In this case, $C^\prime$ is zero except in contiguous troughs which are at least 1000~km~s$^{-1}$ wide and which fall 10\% or more below the emission model; in these troughs $C^\prime \equiv 1$.  The upper integration limit of 29,000~km~s$^{-1}$ is chosen to allow the maximal range of \ion{C}{4} outflow velocities that does not include the \ion{Si}{4}~$\lambda$1400 emission line \citep{thrrsvkafbkn06}.

\begin{figure} [ht]
  \begin{center}
      \includegraphics[width=4in, angle=-90]{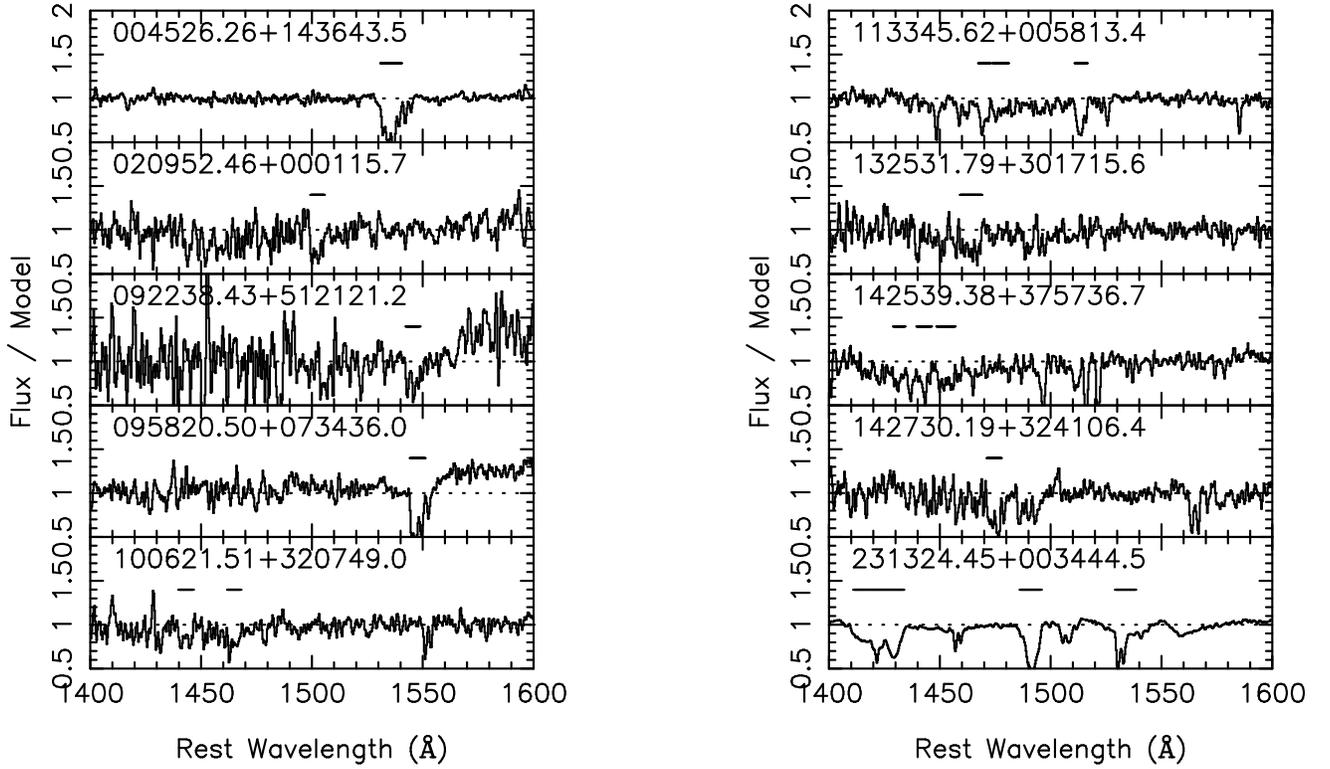}
      \caption{\label{rBTPlotMiniBALExamplesFig}Spectra of ten randomly-selected mini-BAL sources which we classify to have \ion{C}{4} $AI > 0$ and $BI_0 = 0$.  Each spectrum has been divided by a model of the continuum and emission lines.  The asymmetric ``red wing'' of the \ion{C}{4} emission is not modeled, so some cases show an excess at $\lambda \ga 1550$.  The horizontal lines above each spectrum represent regions identified as having mini-BAL absorption.  The spectra have been smoothed by a boxcar of width 3.  A horizontal dotted line indicates where the spectrum-to-model ratio is 1.  See \S\ref{optUVDataSec} for a brief discussion of J$231324.45+003444.5$.}
   \end{center}
\end{figure}

We identify BAL QSOs as those QSOs in our sample with \ion{C}{4} $BI_0 > 0$.  Because we have removed LoBAL sources, the 42 radio-quiet QSOs with $BI_0 > 0$ are all HiBAL QSOs.  We use the $AI$ to formally identify mini-BAL QSOs, although we note that definitions of mini-BALs vary in the literature and some studies consider even narrower features to be mini-BALs.  Requiring \ion{C}{4} $BI_0 = 0$ and $AI > 0$ identifies 48 radio-quiet mini-BAL QSOs with broad \ion{C}{4} absorption features between 1000 and 2000~km~s$^{-1}$ wide.  Randomly-selected examples of mini-BALs in our sample are shown in Figure~\ref{rBTPlotMiniBALExamplesFig}.  Because the defintion of $AI$ extends to higher outflow velocities than that of $BI_0$, it is possible that high-velocity BALs may be mis-classified as mini-BALs.  Inspection of our sources indicates that this is a potential issue only in the single case of J$231324.45+003444.5$, and that this source has additional narrower, mini-BAL features at lower velocities.  We classify this ambiguous source as a mini-BAL for our purposes but do not draw strong conclusions from its individual properties.  Our sample of 90 BAL and mini-BAL QSOs is not large enough to reliably test for bimodality in the distribution of $AI$, as was found in a much larger sample by \citet{ksgc08}.  The remaining 166 sources, with \ion{C}{4} $BI_0 = 0$ and $AI = 0$, are classified as non-BAL QSOs.  The UV and \mbox{X-ray} properties of our radio-quiet sources are described in Table~\ref{uVXRayRQPropertiesTab}.

\subsection{X-Ray Data Reduction\label{xRayDataRedSec}}

Many SDSS QSOs have been observed in targeted or (typically) serendipitous observations with {\it Chandra} or {\it XMM-Newton}.  As part of a previous study to determine the \mbox{X-ray} properties of {\it non}-BAL QSOs, we implemented semi-automated processes to identify such sources and obtain \mbox{X-ray} flux densities from data available in the {\it Chandra} and {\it XMM-Newton} archives.  These procedures are described in detail in \citet{gbs08}.  Briefly, we determined which SDSS QSOs fell on {\it Chandra} ACIS or {\it XMM-Newton} MOS/$pn$ CCDs and reduced these data using standard methods to obtain source and background spectra in each case.  We then used the Cash statistic to fit a broken power law to each unbinned spectrum, with the power law break set at (rest-frame) 2~keV.  From this fit, we calculated the flux density at 2~keV, $F_{\nu}(2~{\rm keV})$, and applied a correction for Galactic absorption.  We determined upper and lower limits on $F_{\nu}$ by adjusting the power law normalization and re-fitting the remaining spectral parameters until the Cash statistic $C$ changed by $\Delta C = 1$.

We used Poisson statistics to determine whether a source was detected (based on background count rates) at $>$99\% confidence in the observed-frame full (0.5--8~keV), soft (0.5--2~keV) or hard (2--8~keV) bands.  High angular resolution is essential to ``resolve away'' the \mbox{X-ray} background and maximize the fraction of source detections.  The excellent resolution of the {\it Chandra} ACIS instrument allowed a high source detection rate even for short exposure times and large off-axis angles.  For this reason, in cases where a source was observed multiple times, we selected (in an unbiased way) the longest {\it Chandra} ACIS exposure as most representative of source properties.  If {\it Chandra} ACIS observations were not available, we selected the longest {\it XMM-Newton} MOS camera observation to determine \mbox{X-ray} fluxes.  We prefer the MOS to the $pn$ camera because the MOS is more effective at ``resolving away'' the background, resulting in a superior detection fraction for these faint sources.

In this work, we consider only sources that are within 10$\arcmin$ of the \mbox{X-ray} observation aim point.  At larger off-axis angles, the angular resolution is significantly worse, leading to greater uncertainty in background estimation and a larger fraction of non-detections.  The large majority of our sources were not specifically targeted in \mbox{X-ray} observations; of the 256 radio-quiet sources in our sample, 222 lie at off-axis angles $>$1\arcmin.

\section{DISCUSSION OF PHYSICAL PROPERTIES\label{analysisSec}}

\subsection{Relative X-Ray Brightness\label{aOXDefnSec}}

We use the parameter $\alpha_{OX}$ to characterize the relation between the observed UV and \mbox{X-ray} luminosities of our sources.  $\alpha_{OX}$ is defined as:
\begin{eqnarray}
\alpha_{OX} &\equiv& 0.3838 \log\biggl( \frac{L_{2~keV}}{L_{2500\mathring{A}}} \biggr),\label{aOXDefnEqn}
\end{eqnarray}
where $L_{2500\mathring{A}}$ and $L_{2~keV}$ are the monochromatic luminosities at 2500~\AA\ and 2~keV, respectively.

We use Equation~3 from \citet{jbssscg07} to determine the typical value of $\alpha_{OX}$ expected for a non-BAL QSO with a UV luminosity of $L_{2500\mathring{A}}$:
\begin{eqnarray}
\alpha_{OX}(L_{2500\mathring{A}}) &=& -0.14 \log(L_{2500\mathring{A}}) + 2.71.\label{justAOXEqn}
\end{eqnarray}
This equation represents an empirical fit to the observed trend for more UV-luminous QSOs to have lower \mbox{X-ray/UV} luminosity ratios \citep[e.g.,][and references therein]{jbssscg07}.

Finally, we quantify the {\it relative} \mbox{X-ray} brightness of a source with respect to a typical non-BAL QSO of the same UV luminosity using the parameter $\Delta\alpha_{OX}$, defined as:
\begin{eqnarray}
\Delta\alpha_{OX} &\equiv& \alpha_{OX} - \alpha_{OX}(L_{2500\mathring{A}}).\label{dAOXDefnEqn}
\end{eqnarray}
$\Delta\alpha_{OX}$ characterizes relative \mbox{X-ray} brightness on a logarithmic scale, so that $\Delta\alpha_{OX} = 0$ indicates a source has the same \mbox{X-ray} luminosity as a typical QSO of the same UV luminosity, while $\Delta\alpha_{OX} = -0.4$ and $-1$ indicate \mbox{X-ray} weakness by factors of 11 and 403, respectively.

\begin{figure} [ht]
  \begin{center}
      \includegraphics[width=5in, angle=-90]{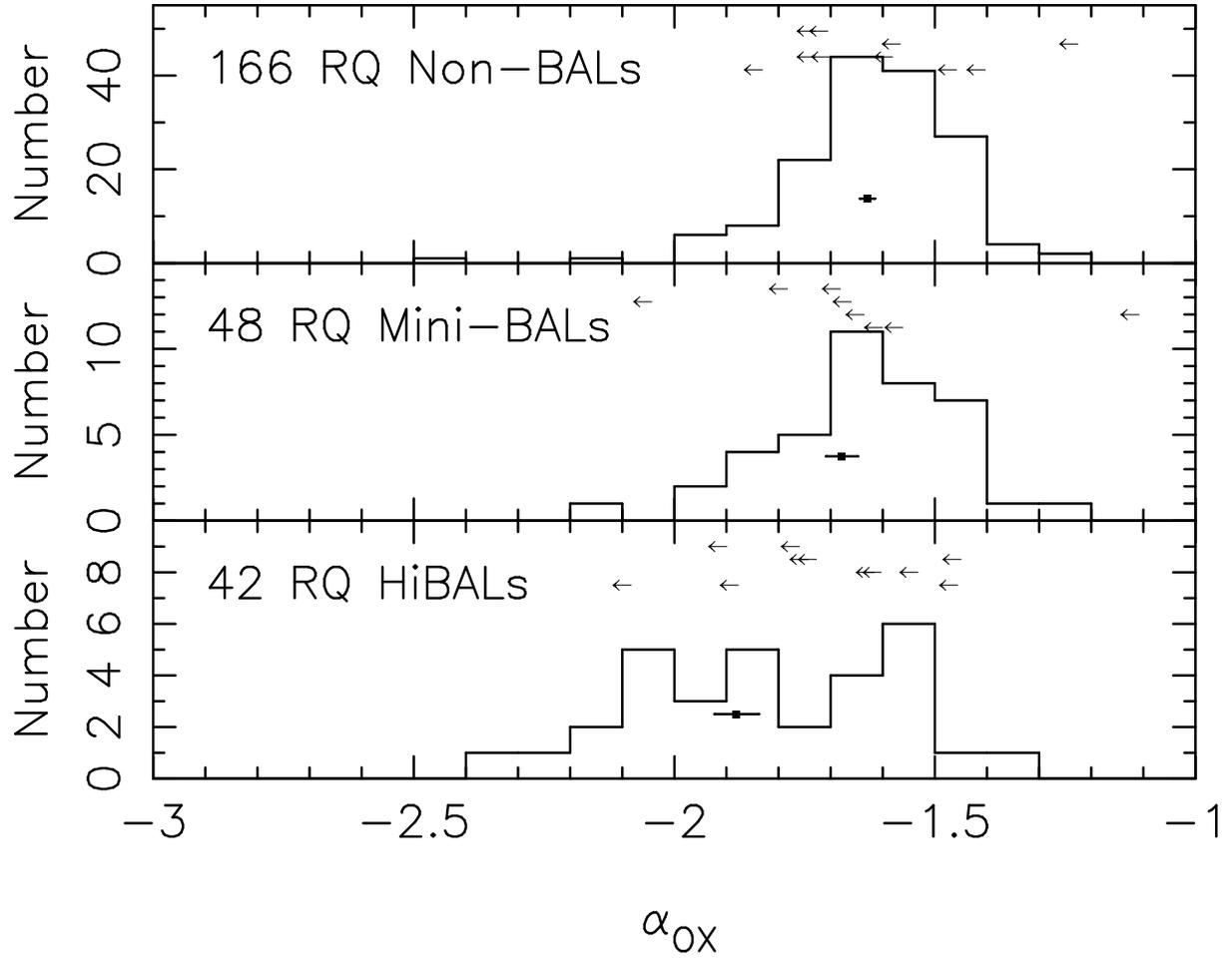}
      \caption{\label{plotRBTAOXFig}The distribution of $\alpha_{OX}$ measured for radio-quiet non-BAL ({\it top}), mini-BAL ({\it middle}), and HiBAL ({\it bottom}) QSOs.  Arrows correspond to 1$\sigma$ upper limits for undetected sources; the $y$-coordinates of upper limit arrows are arbitrary.  The filled black square and its horizontal error bar in each panel represent the mean value of $\alpha_{OX}$ described in \S\ref{aOXDefnSec}.}
   \end{center}
\end{figure}

\begin{figure} [ht]
  \begin{center}
      \includegraphics[width=5in, angle=-90]{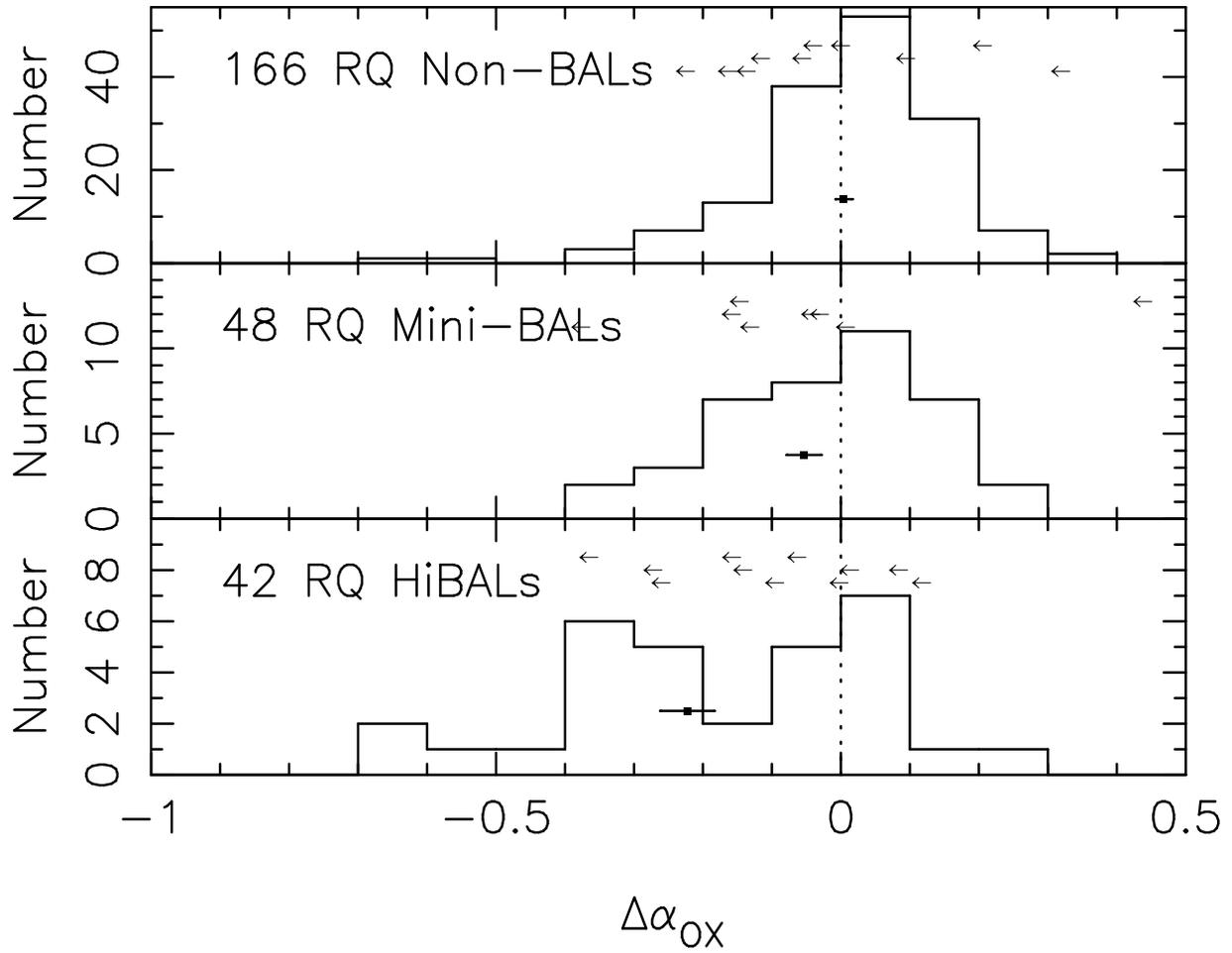}
      \caption{\label{plotRBTDAOXFig}Same as Figure~\ref{plotRBTAOXFig}, but showing the distribution of $\Delta\alpha_{OX}$.  The filled black square and its horizontal error bar in each panel represent the mean value of $\Delta\alpha_{OX}$ described in \S\ref{aOXDefnSec}.}
   \end{center}
\end{figure}

Figure~\ref{plotRBTAOXFig} and Figure~\ref{plotRBTDAOXFig} show the distributions of $\alpha_{OX}$ and $\Delta\alpha_{OX}$, respectively, for the radio-quiet non-BAL, mini-BAL, and HiBAL QSOs in our sample.  In cases where the \mbox{X-ray} emission from a source was not formally detected, we plot an arrow at the value of $\alpha_{OX}$ or $\Delta\alpha_{OX}$ corresponding to our 1$\sigma$ upper limit on $L_{2~keV}$.  Figure~\ref{plotRBTDAOXFig} shows that Equation~\ref{justAOXEqn} describes our non-BAL sample well, as $\Delta\alpha_{OX}$ values for these QSOs are relatively evenly distributed around zero.  The breadth of the non-BAL distribution is attributable largely, if not entirely, to the intrinsic variability of the sources \citep{gbs08}.  HiBAL QSOs, on the other hand, are \mbox{X-ray} weaker than the non-BAL population, with a broad range of $\Delta\alpha_{OX}$ values that are $< 0$ in most cases.  A Gehan test, implemented in the Astronomy Survival Analysis (ASURV) software package \citep[e.g.,][]{if90, lif92}, indicates that the distributions for mini-BAL and non-BAL QSOs differ at 95\% confidence, while the distributions for mini-BAL and HiBAL QSOs differ at 99.7\% confidence. This test accounts for upper limits for sources that were not detected.

The statistical properties of our sources are listed in Table~\ref{sampleLumPropertiesTab}, including the medians and means of $\alpha_{OX}$ and $\Delta\alpha_{OX}$.  The median and mean values were calculated using the Kaplan-Meier estimator implemented in ASURV, which accounts for upper limits on undetected sources.  We find mean $\Delta\alpha_{OX}$ values of $-0.00 \pm 0.01$, $-0.05 \pm 0.03$, and $-0.22 \pm 0.04$ for non-BAL, mini-BAL, and BAL QSOs, respectively.  These values quantitatively demonstrate a trend for mini-BAL QSOs to be slightly \mbox{X-ray} weak compared to non-BAL QSOs; BAL QSOs are generally \mbox{X-ray} weak compared to both mini-BAL and non-BAL QSOs.

Two objects in Figure~\ref{plotRBTDAOXFig} are classified as \mbox{X-ray} detected non-BAL QSOs, yet are quite \mbox{X-ray} weak, with $\Delta\alpha_{OX} < -0.4$.  One of these sources, J$112045.15+130405.2$, is extremely reddened in the UV.  Because the SDSS bandpass does not extend beyond 1400~\AA\ (rest-frame), the continuum placement in the (putative) \ion{C}{4} BAL region is uncertain; this object could be a misclassified BAL QSO.  The other source, J$152156.48+520238.4$, has been analyzed in detail by \citet{jbssscg07}, who argue from the observed UV absorption and hint of a hard \mbox{X-ray} spectrum that the unusual \mbox{X-ray} weakness of this source is likely due to absorbing material along the line of sight, although J$152156.48+520238.4$ is not formally a BAL or mini-BAL QSO.

\subsection{Hard \mbox{X-ray} Photon Indices\label{hardXGammasSec}}

At the redshifts of our sources, hard \mbox{X-rays} are shifted into the more sensitive regions of the {\it Chandra} and {\it XMM-Newton} bandpasses, improving our ability to constrain hard \mbox{X-ray} spectral properties.  In this section, we examine the overall shape of the hard \mbox{X-ray} spectrum, parameterized by the effective photon index $\Gamma$, to determine how it is affected by absorption such as that associated with \mbox{X-ray} weakness in BAL QSOs.

\begin{figure} [ht]
  \begin{center}
      \includegraphics[width=5in, angle=-90]{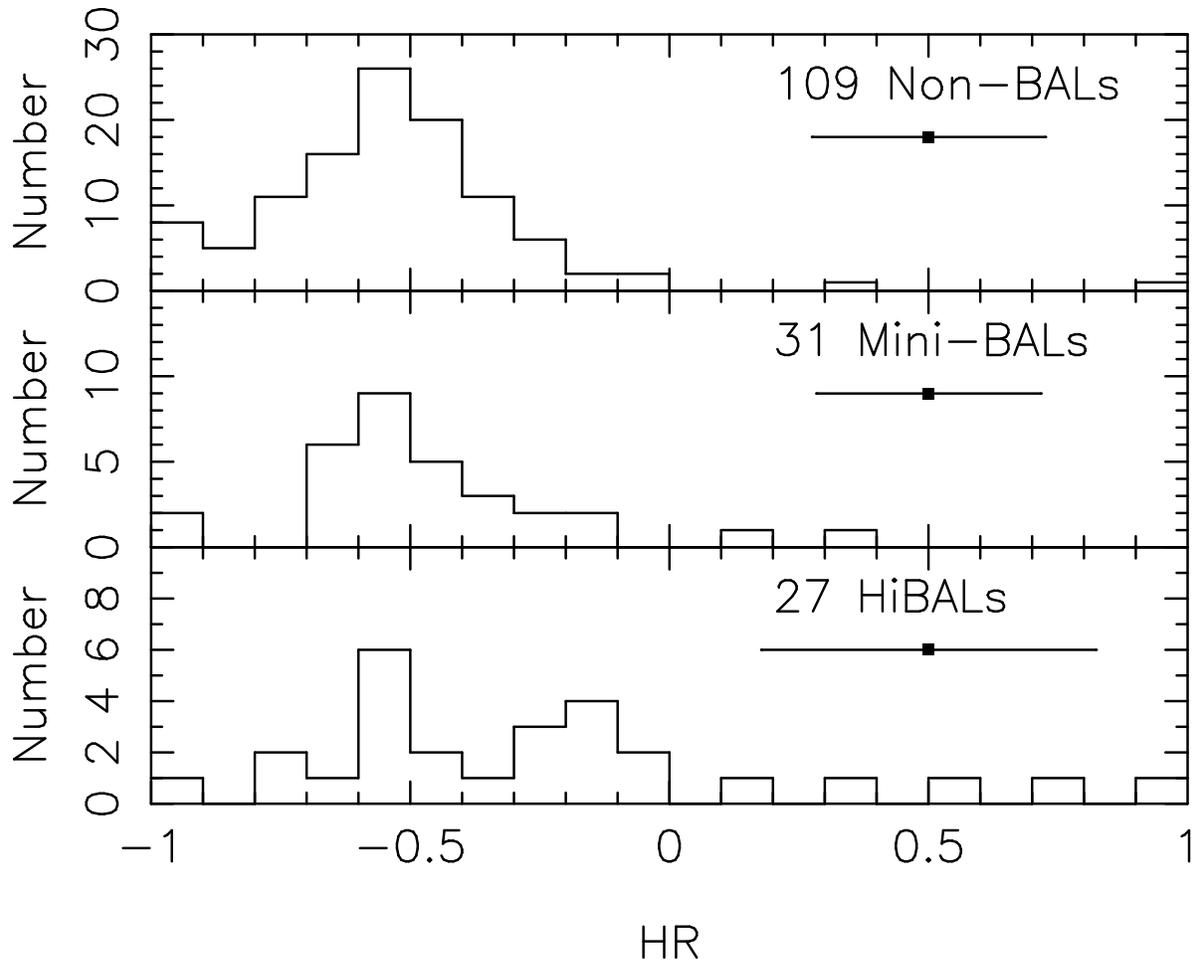}
      \caption{\label{rBTPlotHRChDetHistFig}The distribution of hardness ratios for radio-quiet non-BAL ({\it top}), mini-BAL ({\it middle}), and HiBAL ({\it bottom}) QSOs observed with {\it Chandra}.  The solid point and horizontal lines indicate typical error bars computed formally.}
   \end{center}
\end{figure}

Figure~\ref{rBTPlotHRChDetHistFig} shows \mbox{X-ray} hardness ratios for sources detected with {\it Chandra}.  The detection fraction for sources observed with {\it Chandra} is high, with all 109 non-BAL, 31 (of 34) mini-BAL, and 27 (of 31) BAL QSOs having {\it Chandra} detections.\footnote{The fraction of sources observed with {\it Chandra} is much greater than 50\% because we preferentially select {\it Chandra} observations in cases where both {\it XMM-Newton} and {\it Chandra} have observed a source.}  Limiting the observations to those obtained with {\it Chandra} increases the detection fraction in an unbiased way.  The distributions shown in Figure~\ref{rBTPlotHRChDetHistFig} are therefore reasonably representative of our full sample.  The hardness ratio, $HR \equiv (H-S)/(H+S)$ is calculated from $H$, the number of \mbox{X-ray} counts with observed-frame energies of 2--8~keV, and $S$, the number of counts with observed-frame energies in the 0.5--2~keV range.  The observed hardness ratios suggest that BAL QSOs tend to have flatter spectra than non-BAL and mini-BAL QSOs.  However, hardness ratios do not adequately represent all the information available in the hard \mbox{X-ray} spectrum.  They are also affected by the fact that our sources occur at a range of redshifts and Galactic absorption levels, although this range is relatively small ($1.68 \le z \le 2.28$, $0.01 \le E(B-V) \le 0.11$).

In order to characterize the hard \mbox{X-ray} spectrum as accurately as possible, we calculate the best fit value and 1$\sigma$ confidence limits on the hard \mbox{X-ray} photon index $\Gamma$ for each source in our sample using the broken power law model described in \S\ref{xRayDataRedSec}.  Unlike the hardness ratio calculations, which separate hard and soft \mbox{X-ray} bands at 2~keV in the {\it observed} frame, we in this case define the ``hard power law'' to represent energies $>$2~keV in the {\it rest} frame.  There are not generally enough photons at rest-frame energies $<$2~keV to adequately constrain the soft \mbox{X-ray} spectral shape, so we work with the hard band exclusively.  The values of $\Gamma$ we obtain are intended to constrain the overall hard \mbox{X-ray} spectral shape for sources which may not have sufficient counts to perform a more detailed model fit.  However, we caution that we have measured an effective value of $\Gamma$ that may differ from intrinsic power law photon indices measured with more complex models that include, e.g., the effects of physical absorption models.

\begin{figure} [ht]
  \begin{center}
      \includegraphics[width=5in, angle=-90]{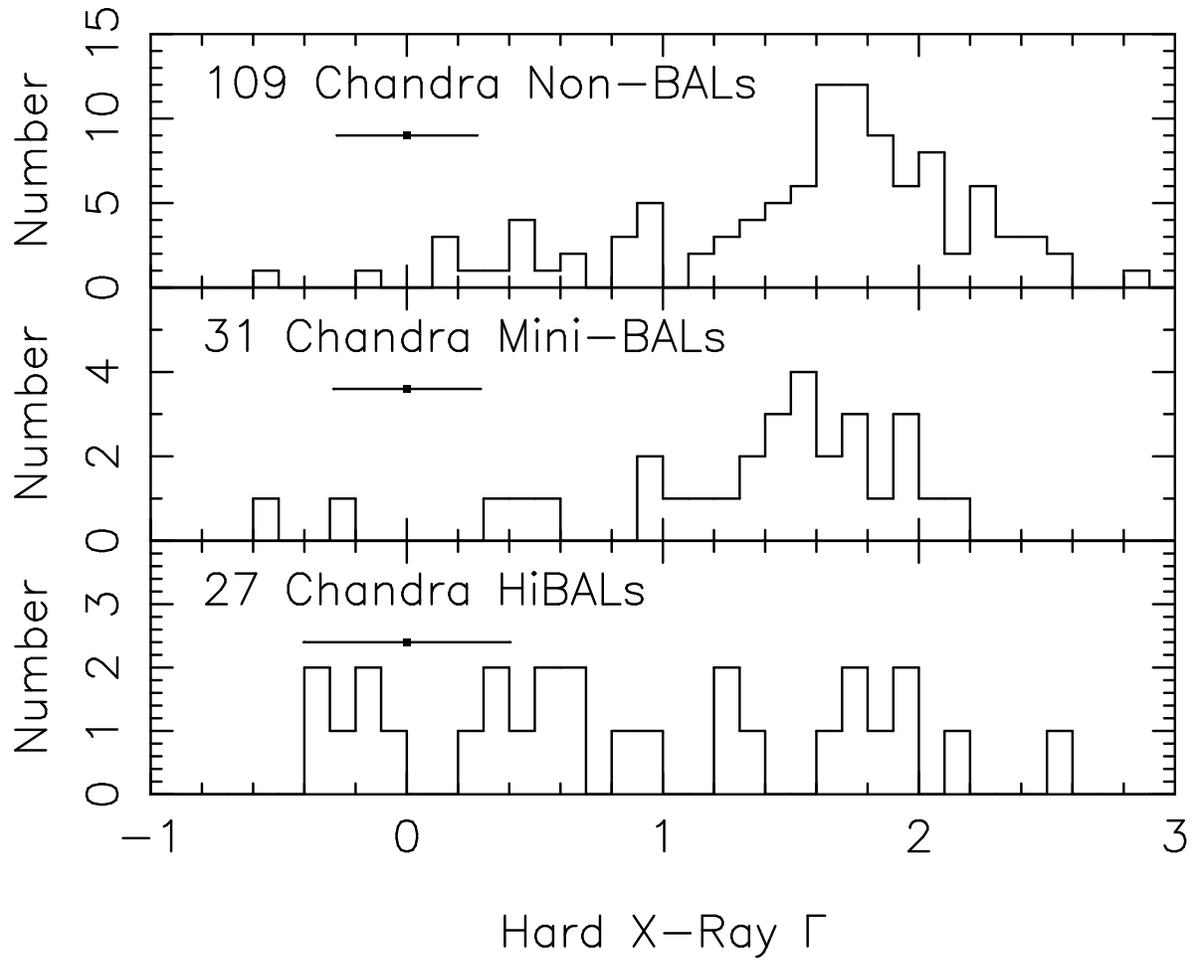}
      \caption{\label{rBTPlotHardXGammasDetHistFig}The distribution of the best-fit hard \mbox{X-ray} photon indices ($\Gamma$) for radio-quiet non-BAL ({\it top}), mini-BAL ({\it middle}), and HiBAL ({\it bottom}) QSOs.    The solid point and horizontal lines indicate typical 1$\sigma$ limits.}
   \end{center}
\end{figure}

The distributions of $\Gamma$ are shown in Figure~\ref{rBTPlotHardXGammasDetHistFig} for radio-quiet non-BAL (top), mini-BAL (middle), and HiBAL QSOs (bottom) which were observed with {\it Chandra} and were detected in \mbox{X-rays}.  Median and mean values of $\Gamma$ are given for these samples in Table~\ref{sampleGammaPropertiesTab}.  A Kolmogorov-Smirnov test indicates that the distribution of $\Gamma$ differs between detected non-BAL and mini-BAL QSOs at 93\% confidence, and between mini-BAL and BAL QSOs at 99.3\% confidence.  We find mean $\Gamma$ values of $1.59 \pm 0.08$, $1.46 \pm 0.18$, and $0.87 \pm 0.16$ for detected non-BAL, mini-BAL, and BAL QSOs, respectively.  (Note that these calculations are affected by outliers, especially at low values of $\Gamma$.)  BAL QSOs have significantly flatter spectra than do non-BAL and mini-BAL QSOs.  Mini-BALs overall have marginally flatter spectra than non-BAL QSOs.

\begin{figure} [ht]
  \begin{center}
      \includegraphics[width=5in, angle=-90]{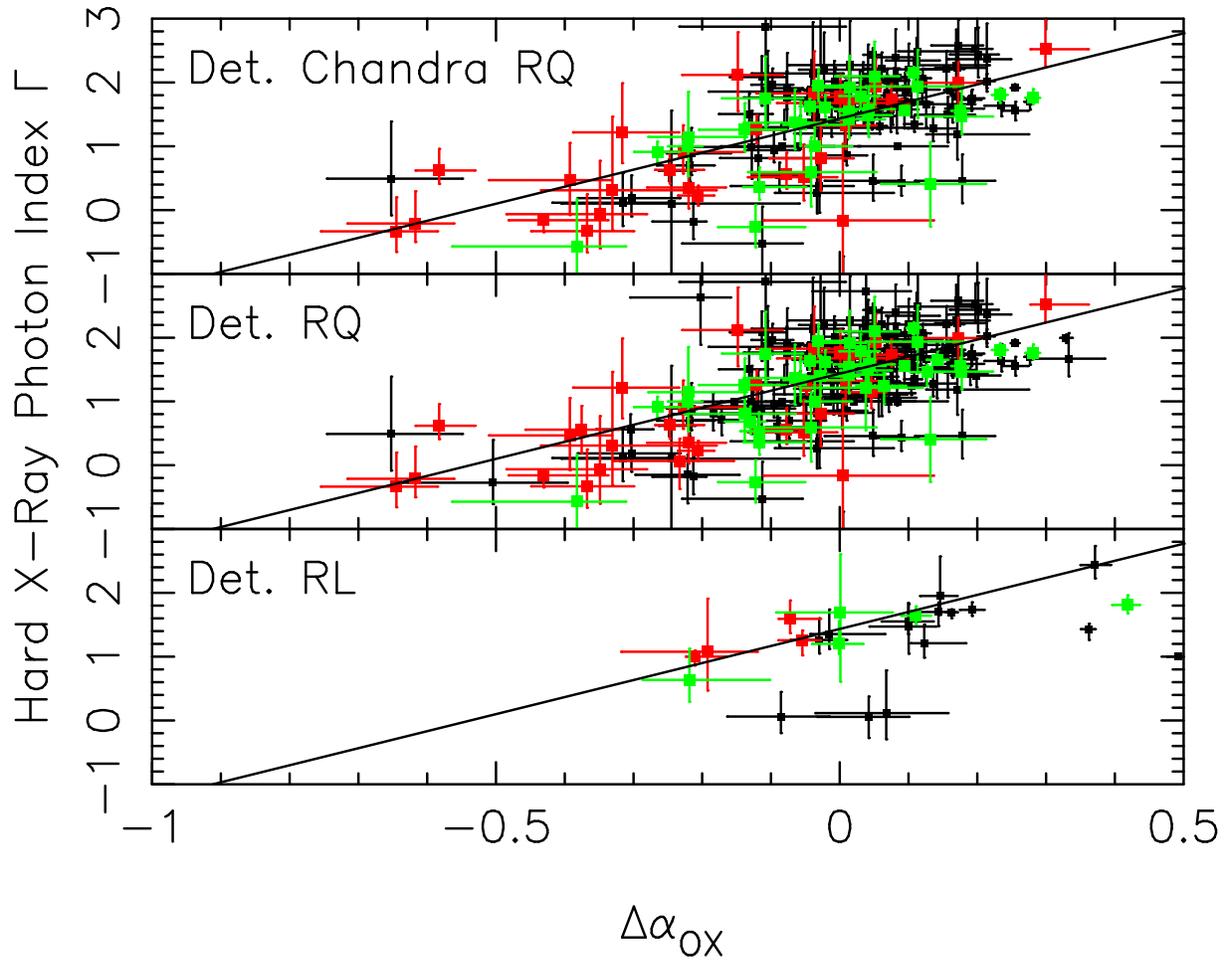}
      \caption{\label{rBTPlotHardXGammasDetFig}The best-fit hard \mbox{X-ray} photon index and 1$\sigma$ confidence ranges for a broken power law model fit to radio-quiet, \mbox{X-ray} detected QSOs plotted against $\Delta\alpha_{OX}$.  Black squares represent non-BAL QSOs, green squares represent mini-BAL QSOs, and red squares represent HiBAL QSOs.  Equation~\ref{rBTPlotHardXGammasDetChandraEqn} is shown as a solid line.}
   \end{center}
\end{figure}

In Figure~\ref{rBTPlotHardXGammasDetFig}, we plot $\Gamma$ against $\Delta\alpha_{OX}$ for radio-quiet, \mbox{X-ray} detected sources.  We plot non-BAL QSOs in black, mini-BAL QSOs in green, and HiBAL QSOs in red.  In the top panel, we plot only sources observed with {\it Chandra}.  As mentioned above, the detection fraction is high for these sources, and the plot therefore is largely representative of non-BAL, mini-BAL, and BAL QSO properties.  A Spearman rank correlation test confirms that $\Gamma$ and $\Delta\alpha_{OX}$ are highly correlated (at $>$99.99\% confidence).  We have fit the sources detected by {\it Chandra} with a linear function $\Gamma(\Delta\alpha_{OX})$ using the EM (estimate and maximize) regression algorithm \citep{dlr77} implemented in ASURV.  The fit line,
\begin{eqnarray}
\Gamma &=& (2.671 \pm 0.262) \Delta\alpha_{OX} + (1.435 \pm 0.046),\label{rBTPlotHardXGammasDetChandraEqn}
\end{eqnarray}
is plotted in each panel of Figure~\ref{rBTPlotHardXGammasDetFig}.  We have excluded from the fit 3 outliers that have $\Gamma > 3$ (not shown in the plot).

In the center panel of Figure~\ref{rBTPlotHardXGammasDetFig}, we show the $\Gamma$ and $\Delta\alpha_{OX}$ values for all radio-quiet sources which were detected either with {\it Chandra} or {\it XMM-Newton}.  The linear fit in Equation~\ref{rBTPlotHardXGammasDetChandraEqn} is overplotted, and is seen to describe the data points well.  In the bottom panel, we plot the data points for the radio-loud, \mbox{X-ray} detected sources, again with the fit line of Equation~\ref{rBTPlotHardXGammasDetChandraEqn} for comparison.  With one exception, the detected HiBAL and mini-BAL QSOs lie near the fit line of Equation~\ref{rBTPlotHardXGammasDetChandraEqn}, and non-BAL QSOs tend to lie below the fit line.  This agrees with previous findings that radio-loud, non-BAL QSOs are relatively \mbox{X-ray} bright \citep[e.g.,][]{wtgz87, blvsbbwg00}.

In order to test whether our fitting procedures artificially induced a correlation between the power law normalization and $\Gamma$, we used our procedures to fit data simulated from models representing a range of values for normalization and $\Gamma$.  We find no artificially-induced correlation, and therefore conclude that the relation between $\Delta\alpha_{OX}$ and $\Gamma$ is due to physical effects.

Previously, \citet{gbcpgs06} have demonstrated that, for BAL QSOs, \mbox{X-ray} spectral slopes inferred from measured hardness ratios are steeper than would be expected if the observed \mbox{X-ray} weakness in BAL QSOs is caused by a neutral absorber with a covering fraction of 100\%.  Here, we briefly investigate whether more complex absorption models can account for the observed tendency for \mbox{X-ray} spectral flattening with weaker \mbox{X-ray} luminosities for BAL, mini-BAL, and non-BAL QSOs.  To do this, we generated grids of two models --- an ionized absorber and a partially-covering, neutral absorber --- in order to identify regions of parameter space that could reproduce Equation~\ref{rBTPlotHardXGammasDetChandraEqn} in the hard \mbox{X-rays}.  For the ionized absorber, we used the pre-generated XSTAR grid 19c table model.\footnote{Available at the XSTAR web site http://heasarc.nasa.gov/lheasoft/xstar/xstar.html}  We used the XSPEC {\tt pcfabs} for a partial covering model.  Overall, we found that both models could reproduce the trend observed between $\Gamma$ and $\Delta\alpha_{OX}$, assuming that the underlying emission was a power law with a fixed $\Gamma \equiv 1.8$, although large absorbing columns ($N_H \ga 10^{22}$~cm$^{-1}$) were of course required to significantly lower $\Gamma$.  Additional constraints on absorption models could be derived from the soft \mbox{X-ray} spectrum, but we do not have sufficient sensitivity to soft \mbox{X-rays} for our sources at rest-frame energies $<$2~keV, as these are redshifted out of the sensitive regions of the detector bandpass.

\subsection{UV Absorption and \mbox{X-ray} Brightness\label{miniHiBALXPropSec}}

\begin{figure} [ht]
  \begin{center}
      \includegraphics[width=5.5in, angle=-90]{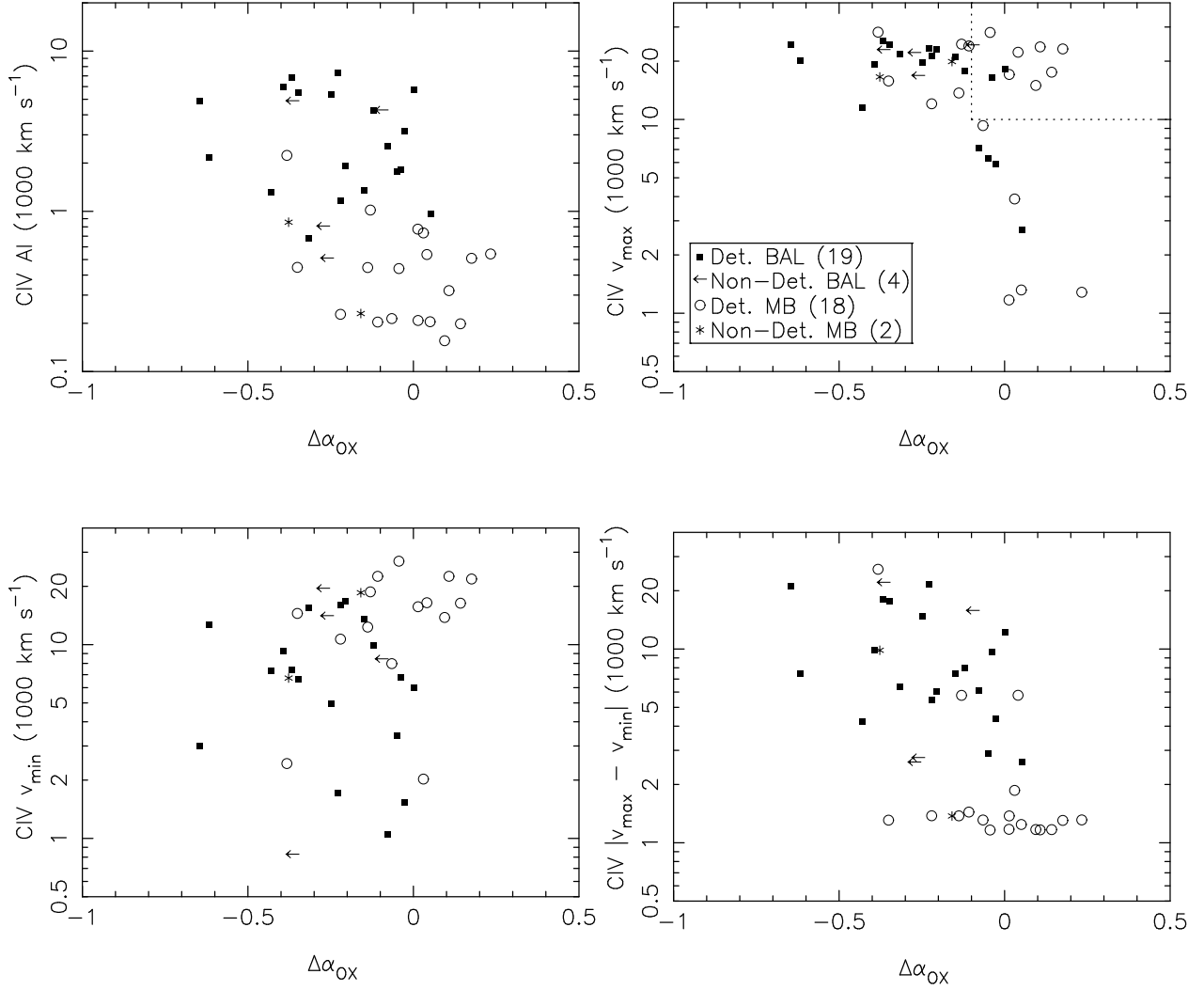}
      \caption{\label{rBTAllVsDAOXFig}{\it Top left:}  the \ion{C}{4} absorption index, $AI$ (Equation~\ref{aIDefnEqn}), plotted against $\Delta\alpha_{OX}$ for radio-quiet HiBAL and mini-BAL QSOs with $SN_{1700} \ge 9$.  BAL QSOs are plotted with filled squares for detected sources or arrows indicating upper limits.  Mini-BALs are represented by open circles for detected sources or asterisks for upper limits.  {\it Top right:}  the \ion{C}{4} BAL or mini-BAL $v_{max}$ against $\Delta\alpha_{OX}$.  The dotted lines indicate the region used to distinguish mini-BAL samples $A$ and $B$, described in \S\ref{miniHiBALXPropSec}.  {\it Bottom left:}  the \ion{C}{4} BAL or mini-BAL $v_{min}$ against $\Delta\alpha_{OX}$.  {\it Bottom right:}  the \ion{C}{4} BAL or mini-BAL $\Delta v \equiv |v_{max} - v_{min}|$ against $\Delta\alpha_{OX}$.  In cases where multiple mini-BAL troughs are present, $\Delta v$ can be $>$2000~km~s$^{-1}$; a small number of such cases are evident in the plot.}
   \end{center}
\end{figure}

In Figure~\ref{rBTAllVsDAOXFig}, we plot UV absorption properties against $\Delta\alpha_{OX}$ for \ion{C}{4} BAL (filled squares) and mini-BAL (open circles) QSOs which are radio-quiet and have $SN_{1700} > 9$, where $SN_{1700}$ is defined as the median of the flux divided by the noise (as reported by the SDSS pipeline) for all spectral bins in the 1650--1700~\AA\ region.  We exclude sources with lower $SN_{1700}$ in order to obtain the best possible measurements of trough velocities and widths for this analysis.  For sources not detected in \mbox{X-rays}, we plot upper limits with arrows (BAL QSOs) or arrows within circles (mini-BAL QSOs).  According to Kendall's Tau test implemented in ASURV, $AI$ and $\Delta\alpha_{OX}$ are correlated at 99.8\% confidence.  Because mini-BALs have lower values of $AI$ than do BAL QSOs, and because they are relatively \mbox{X-ray} bright compared to BAL QSOs (\S\ref{aOXDefnEqn}), they extend the trend for UV absorption strength to decrease with relative \mbox{X-ray} brightness \citep[e.g.,][]{blw00, gbcpgs06, g+08c}.

We define $v_{max}$ as the outflow velocity corresponding to the shortest wavelength which is identified as part of a \ion{C}{4} mini-BAL or BAL trough.  Similarly, $v_{min}$ is the outflow velocity associated with the longest wavelength in a trough.  Figure~\ref{rBTAllVsDAOXFig} shows \ion{C}{4} $v_{max}$ and $v_{min}$ values plotted against $\Delta\alpha_{OX}$ for the same sources.  There is a significant correlation between $v_{max}$ and $\Delta\alpha_{OX}$ (at 99.3\% confidence); if we exclude mini-BALs, the correlation strengthens only slightly (99.7\% confidence).  In the case of $v_{min}$, no significant correlation is found.  We also show the distribution of velocity widths, $\Delta v \equiv |v_{max} - v_{min}|$, as a function of $\Delta\alpha_{OX}$.  The two quantities are correlated at $>$99.99\% confidence.  This is not surprising, given that mini-BAL QSOs have narrower absorption troughs by definition and are observed to be somewhat brighter in \mbox{X-rays} than BAL QSOs are.

Four mini-BAL QSOs have multiple troughs $>$1000~km~s$^{-1}$ wide.  For these sources, $v_{max}$ represents the highest outflow velocity of the higher-velocity trough, and $v_{min}$ represents the lowest outflow velocity of the lower-velocity trough.  With this convention, mini-BAL QSOs with multiple troughs can have $\Delta v > 2000$~km~s$^{-1}$, even though no single trough is that broad.

The correlations are weakened by a population of mini-BAL QSOs with high outflow velocities ($v_{min}$ and $v_{max} \ga 10,000$~km~s$^{-1}$) that have relatively normal observed \mbox{X-ray} luminosities.  Visual inspection of the UV spectra of the high-velocity ($|v_{max}| > 10,000$~km~s$^{-1}$), \mbox{X-ray} normal ($\Delta\alpha_{OX} > -0.1$) mini-BAL QSOs does not reveal any spectral characteristics unique to this population.  To perform a more quantitative study, we divided the mini-BAL sample into two subsamples.  Subsample $A$ contains the 7 radio-quiet mini-BALs with $SN_{1700} > 9$, $|v_{max}| > 10,000$~km~s$^{-1}$, and $\Delta\alpha_{OX} \ge -0.1$.  Subsample $B$ contains the remaining 13 radio-quiet mini-BAL QSOs with $SN_{1700} > 9$.  (These somewhat subjective criteria were chosen in an attempt to identify the high-velocity, \mbox{X-ray} bright population from Figure~\ref{rBTAllVsDAOXFig}.)  A Gehan test, implemented in ASURV, finds no significant difference between the distributions of $L_{\nu}(2500~\mathring{A})$ or \ion{C}{4} $AI$ values for these two subsamples.  \mbox{X-ray} monochromatic luminosities, $L_{\nu}(2~{\rm keV})$, differ at 99\% confidence, but this result is likely an artifact of our selection criteria that sample $A$ be relatively \mbox{X-ray} bright, given that the UV luminosity range of our sample is limited.  There is a mild distinction (at 97\% confidence) in redshift $z$ between subsamples $A$ and $B$, with sample $A$ having generally lower redshifts.  Both systematic effects (involving the calculation of $AI$) and cosmological effects (involving intervening absorbers) could influence this weak relation, so we do not draw any physical conclusion from it.

We note that the four narrower intrinsic absorption systems studied by \citet{mecc08} also had high values of $|v_{max}|$.  If they had broader (mini-BAL) UV absorption, three of their four sources would fall in our sample $A$, and the fourth source would also be included if it were only slightly brighter (by $\Delta\alpha_{OX} \approx 0.05$).  These studies suggest that a significant population of high-velocity NAL and mini-BAL absorbers may be found in relatively \mbox{X-ray} bright QSOs, in apparent contrast to the tendency for high-velocity BALs to appear in relatively \mbox{X-ray} weak QSOs.

\subsection{Pointed XMM Observations\label{ourXMMSec}}

\begin{figure} [ht]
  \begin{center}
      \includegraphics[width=6in, angle=-90]{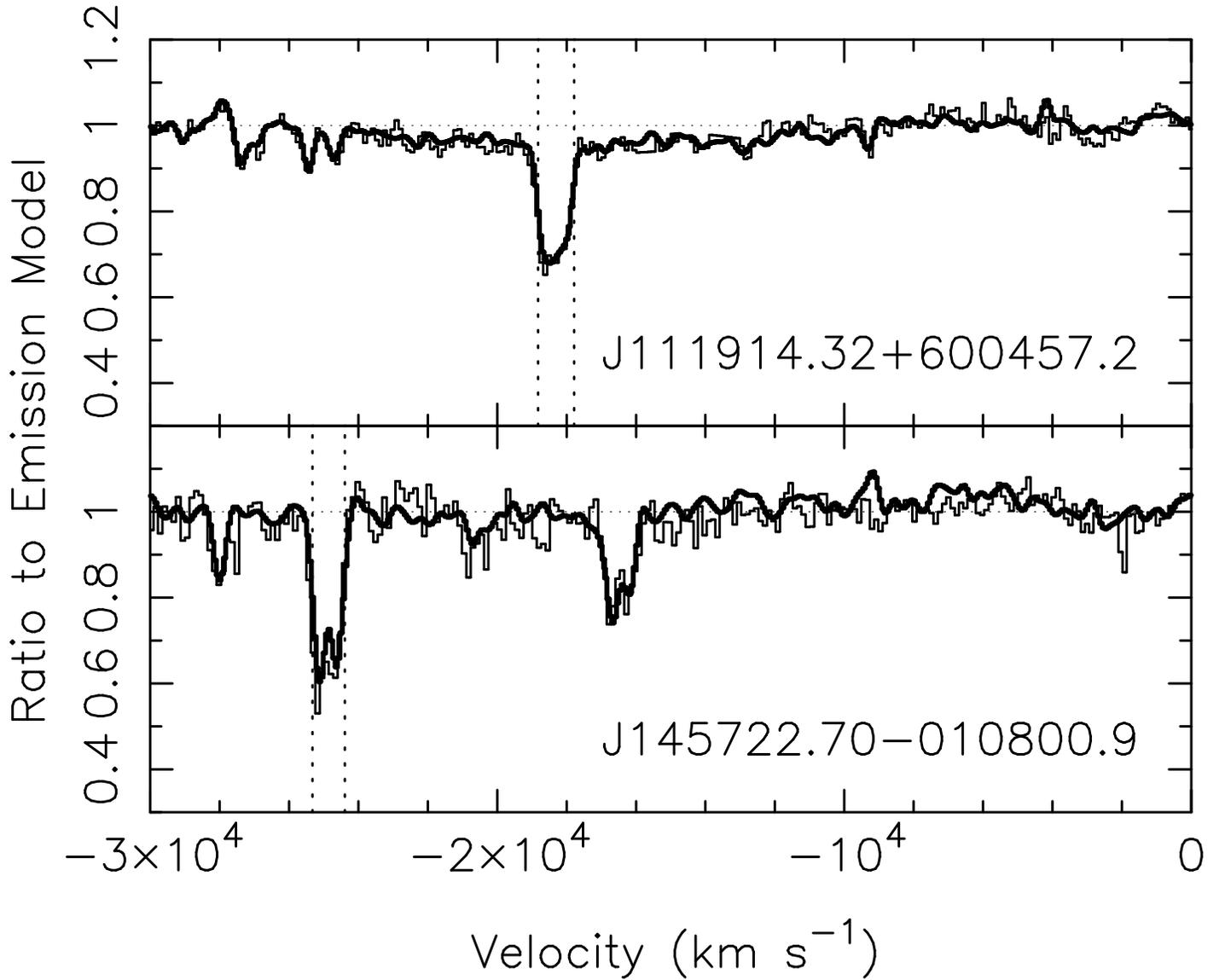}
      \caption{\label{plotRBTUVRatioFig}The SDSS (thick line) and HET (thin line) spectra of SDSS J$111914.32+600457.2$ (top) and J$145722.70-010800.9$ (bottom).  The SDSS spectra (representing an earlier epoch) have been smoothed with a Gaussian to match the resolution of the HET spectra.  Each spectrum has been divided by a fit to the continuum and broad line emission in the region shown.  Zero velocity corresponds to the rest-frame wavelength of the red line in the \ion{C}{4} doublet.  A horizontal dotted line indicates where the $y$-axis is 1.  The vertical dotted lines mark the regions identified as mini-BALs by \citet{thrrsvkafbkn06}.  (The absorption feature at velocity $\approx$--16,000~km~s$^{-1}$ in J$145722.70-010800.9$ is slightly too narrow to formally have $AI > 0$.)  The spectral resolution is $\approx$1.7 and 1.9~\AA\ for the top and bottom panels, respectively.}
   \end{center}
\end{figure}

Following the procedures described in \S\ref{xRayDataRedSec}, we have obtained \mbox{X-ray} luminosities using pointed {\it XMM-Newton} observations of two mini-BAL sources from the \citet{thrrsvkafbkn06} BAL catalog which were identified as unusually \mbox{X-ray} bright based on their {\it ROSAT} count rates \citep{vabbbbdeghhhkppprstz99, vabbbbdeghhhpppstz00}.  The UV and \mbox{X-ray} luminosities, $\alpha_{OX}$, and $\Delta\alpha_{OX}$ values measured for these sources are given in Table~\ref{makeXMMObsTableTab}.  In Figure~\ref{plotRBTUVRatioFig}, we present their SDSS and HET spectra.  For comparison, the SDSS spectra have been smoothed with a Gaussian to approximate the HET spectral resolution.  Vertical, dotted lines indicate the regions identified as mini-BALs by \citet{thrrsvkafbkn06}.  The mini-BAL in J$145722.70-010800.9$ may have strengthened slightly, but otherwise there is no evidence for variation in the spectra over the 1.5 and 1.8~yr (QSO frame) times between observations.  The mini-BAL in J$145722.70-010800.9$ has a doublet structure, while that of J$111914.32+600457.2$ resides in the center of a broad but very shallow trough (from approximately $-12,000$ to $-22,000$~km~s$^{-1}$) and has no visible doublet structure.

In light of our observation that mini-BAL QSOs are relatively \mbox{X-ray} bright compared to BAL QSOs, our {\it XMM-Newton} observations (Table~\ref{makeXMMObsTableTab}) indicate that these sources are not highly atypical for radio-quiet (J$145722.70-010800.9$) or radio-loud (J$111914.32+600457.2$) mini-BAL QSOs.  In order to compare our \mbox{X-ray} brightness measurements with those predicted from the earlier {\it ROSAT} observations, we performed manual photometry on the {\it XMM-Newton} and {\it ROSAT} images and used PIMMS,\footnote{http://heasarc.gsfc.nasa.gov/Tools/w3pimms.html} to test for evidence of source variability.

For J$145722.70-010800.9$, a source ($145723-010723$) which is about 70\% brighter is located nearby in the {\it XMM-Newton} \mbox{X-ray} image.  $145723-010723$ is likely related to a galaxy near that location identified in the SDSS database.  The SDSS estimated photometric redshift for this galaxy is $z = 0.44 \pm 0.07$.  The doublet absorption features in the SDSS spectrum of J$145722.70-010800.9$ do not match lines expected for an intervening absorber at $z \approx 0.44$.

The source $145723-010723$ is about 24\arcsec, or 2$R_{\sigma}$ from the {\it ROSAT} catalog source identification, where $R_{\sigma} = 12$\arcsec~is the {\it ROSAT} $1\sigma$ positional error (including 6\arcsec of systematic error).  J$145722.70-010800.9$ is somewhat closer ($1.2 R_{\sigma}$) to the {\it ROSAT} source identification \citep{vabbbbdeghhhpppstz00}.  The 60\arcsec~radius region centered on J$145722.70-010800.9$ in the {\it ROSAT} image blends the two sources and accounts for about half of the count rate reported in the Faint Source Catalog based on that catalog's extraction radius of 300\arcsec.  The {\it ROSAT} count rate is 3--4 times higher than expected from estimates based on the two {\it XMM-Newton} sources combined, suggesting that some source in this region was brighter in the earlier epoch.  Because the {\it ROSAT}-based identification was closer to J$145722.70-010800.9$, it seems possible that this mini-BAL QSO was brighter in the earlier epoch.

The situation is more complicated for J$111914.32+600457.2$, as the excess {\it ROSAT} counts over the background are not as centrally peaked as was the case for J$145722.70-010800.9$, making the exact count rate estimate highly dependent on aperture size.  Using a 60\arcsec~radius aperture, we estimate that the source J$111914.32+600457.2$ has dimmed by $\sim$50\%, but this factor could be up to $\sim$90\% with an aperture as large as 300\arcsec~in radius.  Our {\it XMM-Newton} images show two somewhat fainter sources $\approx$200\arcsec~distant from J$111914.32+600457.2$, but these sources do not coincide with the brightest regions of the 300\arcsec~{\it ROSAT} aperture.  As for J$145722.70-010800.9$, we conclude that some variable source contributed to the excess of counts in the {\it ROSAT} epoch, but we cannot be certain that the variable source was the QSO $111914.32+600457.2$.

\section{DISCUSSION\label{discSec}}

\subsection{Are Mini-BALs Just Narrow BALs?\label{areMBsIntermSec}}

Our analysis of the \mbox{X-ray} properties of mini-BAL QSOs indicates that the relative \mbox{X-ray} brightness and hard \mbox{X-ray} spectral slopes of mini-BAL QSOs are, in general, intermediate between those of non-BAL and BAL QSOs.  This supports previous suggestions that at least some narrower and broader absorption phenomena are physically related \citep[e.g.,][]{akdjb99, gbcg02, gb08, mecc08}.  Flatter hard \mbox{X-ray} spectra observed for QSOs with mini-BAL and especially BAL absorption could be explained by ionized or partial-covering absorber models with high ($N_H \ga 10^{22}$~cm$^{-2}$) effective column densities.

However, we caution that classifying mini-BAL QSOs entirely as ``intermediate'' between BAL and non-BAL QSOs is a general statement that does not address the full complexity of absorbing outfows with intermediate velocity widths.  Averaged mini-BAL QSO properties, such as relative \mbox{X-ray} brightness and hard \mbox{X-ray} $\Gamma$, are closer to those of non-BAL QSOs than they are to those of BAL QSOs.  Some mini-BAL QSOs do not follow the trends previously observed for BAL QSOs between outflow velocity and relative \mbox{X-ray} brightness $\Delta\alpha_{OX}$.  The four sources with narrower intrinsic absorption in the study of \citet{mecc08} were also found to be relatively \mbox{X-ray} bright and to have high outflow velocities compared to BAL QSOs.

Furthermore, we note that the technical definition of ``mini-BAL'' encompases a wide range of absorption features.  For example, troughs with a doublet-like structure can be classified as mini-BALs.  Weak, narrow absorption features that lie inside broad absorption troughs that are themselves too shallow to be classified as BALs may also be classified as mini-BALs.  Absorption features caused by \ion{Al}{1}~$\lambda$1414 and \ion{S}{1}~$\lambda$1425 could, in principle, be mis-categorized as high-velocity \ion{C}{4} mini-BALs, although this identification would require highly non-solar elemental abundances to explain the absence of, e.g., \ion{C}{1}~$\lambda$1657 absorption.  It has been noted that the distribution of the absorption index, $AI$, is bimodal \citep{ksgc08}, suggesting that (at least) two different varieties of absorption phenomena may be combined in the BAL and mini-BAL categories.  As previously mentioned, visual inspection of UV absorption does not indicate any notable differences between the high-velocity, relatively \mbox{X-ray} bright mini-BAL QSOs in our Sample~A and those of Sample~B, which have relative \mbox{X-ray} brightness more typical of BAL QSOs.  However, further study of the characteristics of mini-BAL troughs is required to determine whether the mathematical definition of ``mini-BAL'' is adequately representing the wide range of observed UV absorption phenomena.

One may speculate that the mini-BALs of sample $A$ (with $|v_{max}| > 10,000$~km~s$^{-1}$ and $\Delta\alpha_{OX} > -0.1$) are caused by intervening, rather than intrinsic, absorbers.  While we cannot conclusively rule out the possibility of intervening systems contaminating our sample, we note that the velocity widths ($>$1000~km~s$^{-1}$) required for mini-BALs would require a high spread of velocities in any intervening absorber sufficient to broaden the absorption feature (which has doublet lines spaced $\approx$500~km~s$^{-1}$ apart) by $\approx$500~km~s$^{-1}$.  In general, mini-BALs have also been observed to vary more commonly than do NALs \citep[e.g.,][and references therein]{nhbbcjl04}, and several of our results are consistent with those of \citet{mecc08}, which were obtained for a small sample of narrow absorption lines which were known to be variable.  Variability is commonly taken to indicate that the absorber is relatively compact and therefore likely to be intrinsic.  Future optical/UV observations may indicate whether the absorption features of QSOs in sample $A$ are variable.

\subsection{Mini-BAL and BAL Absorbers\label{mBBALAbsorbersSec}}

\citet{mecc08} have adopted a model from \citet{gbcebc01} in which the NAL material is located outside the equatorial BAL outflow, so that lines of sight through NAL absorbers correspond to smaller inclination angles with respect to the accretion disk normal than do lines of sight through BAL outflows.  The NAL absorbers have smaller absorbing columns than the BAL region does, but they reside sufficiently far from the central source that they are not overionized and can be radiatively accelerated to high speeds.

Alternatively, we propose a simple model that also takes into account previous claims for a bifurcation in absorber properties \citep[e.g.,][]{ksgc08} as well as our observation that strong \mbox{X-ray} absorption is apparently not always required to accelerate mini-BALs to high outflow velocities.  We have previously suggested that BAL variation in relatively narrow velocity regions might be associated with clumps of absorbing material which are present in the BAL outflow \citep{gbsg08}.  Suppose the radial velocities of these clumps are independent of the acceleration mechanism for BALs.  They may be ``seeds'' of UV-absorbing material, launched by an unknown physical mechanism in the QSO nucleus, which are evaporated over time by the QSO continuum.  If conditions are right, the evaporated material can be radiatively accelerated to form broad BAL troughs, but if the incident \mbox{X-ray} flux is too high, the outflow is overionized and a BAL cannot form.  In that case, a population of high-velocity, \mbox{X-ray} bright mini-BALs could represent failed absorption ``seeds'' which entered the line of sight at a high velocity, but their evaporated material could not be effectively accelerated into an outflow covering a wide velocity range.  With stronger \mbox{X-ray} absorption, the mini-BALs would have been accelerated into broader features, and the UV and \mbox{X-ray} properties would have been similar to those observed for BAL QSOs.

\subsection{The Search for X-Ray Bright BALs\label{searchXBBsSec}}

In \S\ref{ourXMMSec}, we discussed the results of our {\it XMM-Newton} observations of mini-BAL QSOs which were identified as unusually \mbox{X-ray} bright based on their {\it ROSAT} count rates.  In our targeted observations, the sources were measured to have values of $\Delta\alpha_{OX}$ which were not highly atypical when compared to those of mini-BAL QSOs.  The high {\it ROSAT} count rates may be at least partly attributable to source variability.

Our current study identifies two additional HiBAL sources as being relatively \mbox{X-ray} bright (for {\it BAL} QSOs):  J$121205.30+152005.8$ and J$143031.78+322145.9$ (with $\Delta\alpha_{OX} > 0.1$).  Both sources were observed for 5~ks by {\it Chandra} (observation id numbers 2104 and 4279, respectively), yielding \mbox{X-ray} spectra with $\approx$30 counts each.  These spectra are insufficient for detailed spectral analysis.

Recently, \citet{gp08} have identified three additional candidates as potential \mbox{X-ray} bright BAL QSOs.  Applying our classification methods to their sources, we identify one of their sources, J$1019442.92+450239.4$, as a LoBAL QSO with \ion{Si}{4}, \ion{C}{4}, and \ion{Al}{3} BALs evident.  We classify their source J$023219.52+002106.8$ as having a mini-BAL but not a BAL.  Their third source, J$132228.37+141022.8$ is not classified as either a BAL or a mini-BAL QSO; the identification of broad \ion{C}{4} absorption in this source requires spectral smoothing with a broader window than the 3 bins used in our methods.

Two recent studies have examined samples of BAL QSOs which were selected due to their detection in \mbox{X-rays}.  Because \mbox{X-ray} brightness was a criterion in defining their samples, these studies identified relatively \mbox{X-ray}-bright BAL QSOs.   \citet{gcv08} found that BAL QSOs detected by {\it XMM-Newton} tended to have lower \mbox{X-ray} absorbing columns than did optically-selected BAL QSOs, and the $\alpha_{OX}$ values of these sources were similar to those of non-BAL QSOs.  In general, they found that sources with weaker UV absorption features tended to be less absorbed in \mbox{X-rays}.  \citet{gcv08} note that some of their sources show relatively weak UV absorption that may not be classified as BALs using the formal $BI > 0$ criterion.  On the other hand, \citet{bdpmsklms08} found significant \mbox{X-ray} absorption in their sample of 5 \mbox{X-ray}-selected BAL QSOs with stronger UV absorption, and hypothesized that their sources exhibited relatively high \mbox{X-ray}-to-optical intrinsic flux ratios.  While these studies have a somewhat different focus from our current study, \citet{gcv08} agree with our observation that sources with weaker UV absorption (i.e., mini-BALs) are in general (though not always) relatively \mbox{X-ray} brighter than BAL QSOs are, and they demonstrate the need for additional study of the relation between \mbox{X-ray} and UV absorption.

On the basis of our current study, we suggest several criteria that should be evaluated in order to appropriately confirm cases of \mbox{X-ray} bright BAL QSOs.  First, the source should be securely identified at high angular resolution, to confirm that \mbox{X-ray} photons are not contributed from nearby sources.  Second, our study has shown that mini-BAL QSOs are generally \mbox{X-ray} brighter than BAL QSOs.  Therefore, the sources of most interest should be securely identified as {\it bona fide} BAL QSOs, with strong, broad ($>$2000~km~s$^{-1}$) absorption troughs.  Third, the source should be known to be radio-quiet, as radio-loud QSOs are generally \mbox{X-ray} brighter and may have jet-linked \mbox{X-ray} emission exterior to any BAL absorption.  Finally, \mbox{X-ray} and optical/UV spectroscopy would ideally be obtained simultaneously, to account for any possible variation or evolution in the \mbox{X-ray} and BAL absorbers.  Multiple \mbox{X-ray} observations would also be useful to test for emission and absorption variability.

\section{CONCLUSIONS\label{concSec}}

\begin{enumerate}
\item{In general, mini-BAL QSOs are intermediate between BAL and non-BAL QSOs in terms of \mbox{X-ray} brightness and effective hard \mbox{X-ray} photon index, $\Gamma$, as might be expected if at least some narrower and broader absorption phenomena are physically related \citep[e.g.,][]{akdjb99, gbcg02, gb08, mecc08}.}
\item{The steepness of the hard \mbox{X-ray} spectrum, parameterized by an effective power law photon index, $\Gamma$, is correlated with the relative \mbox{X-ray} brightness, $\Delta\alpha_{OX}$, for a combined sample of radio-quiet BAL, mini-BAL, and non-BAL QSOs.}
\item{Because mini-BAL QSOs have weaker UV absorption and are relatively \mbox{X-ray} bright compared to BAL QSOs, they extend previously-observed trends between UV absorption strength and $\Delta\alpha_{OX}$ observed for BAL QSOs.}
\item{However, a significant population of ``\mbox{X-ray} normal'' (with $\Delta\alpha_{OX} \sim 0$) mini-BAL QSOs is found to have high outflow velocities, contrary to observed trends for BAL QSOs.  This could be explained if some mini-BAL absorbers were launched (by an unknown mechanism) with a range of velocities, but could not be radiatively accelerated into broader absorption features due to inadequate \mbox{X-ray} shielding.}
\item{The existence of a population of mini-BAL QSOs with high \ion{C}{4} outflow velocities that are not \mbox{X-ray} weak indicates that consistently strong \mbox{X-ray} absorption is not always required to launch or accelerate at least some mini-BAL absorbers.}
\item{The \mbox{X-ray} brightest QSOs with broad UV absorption are typically mini-BALs.  We have shown that mini-BAL QSOs are, in general, relatively bright in \mbox{X-rays} compared to BAL QSOs.  Follow-up observations at higher angular resolution suggest that previous identifications of anomalously \mbox{X-ray} bright QSOs with broad UV absorption may be at least partly attributable to source variability.}
\end{enumerate}

\acknowledgements
We gratefully acknowledge support from NASA LTSA grant NAG5-13035 (RRG, WNB, DPS), NASA grant NNX07AQ57G (RRG, WNB), the National Science and Engineering Research Council of Canada (SCG), and NSF grant AST06--07634 (DPS).

Funding for the SDSS and SDSS-II has been provided by the Alfred P. Sloan Foundation, the Participating Institutions, the National Science Foundation, the U.S. Department of Energy, the National Aeronautics and Space Administration, the Japanese Monbukagakusho, the Max Planck Society, and the Higher Education Funding Council for England.  The SDSS Web site \hbox{is {\tt http://www.sdss.org/}.}

The Hobby-Eberly Telescope is a joint project of the University of Texas at Austin, the Pennsylvania State University, Stanford University, Ludwig-Maximilians-Universit\"{a}t M\"{u}nchen, and Georg-August-Universit\"{a}t G\"{o}ttingen. The HET is named in honor of its principal benefactors, William P. Hobby and Robert E. Eberly.

The Marcario Low Resolution Spectrograph is named for Mike Marcario of High Lonesome Optics who fabricated several optics for the instrument but died before its completion. The LRS is a joint project of the Hobby-Eberly Telescope partnership and the Instituto de Astronomía de la Universidad Nacional Autónoma de México.



%\appendix{\label{emLineFitAssumptionsSec}}
%\section{Emission Line Fits Assumptions}

% -------------- BEGIN BIBLIOGRAPHY -----------------

\bibliographystyle{apj3}
\bibliography{apj-jour,bibliography}

% -------------- BEGIN TABLES -----------------
\begin{deluxetable}{llllll}
\tabletypesize{\scriptsize}
\tablecolumns{6}
\tablewidth{0pc}
\tablecaption{Observation Log\label{obsLogTab}}
\tablehead{\colhead{SDSS DR5 Name} & \colhead{{\it XMM-Newton} Date} & \colhead{SDSS Date} & \colhead{HET Date}}
\startdata
J$111914.32+600457.2$ & 2007 May 20 & 2002 May 4 & 2007 Nov 21\\
J$145722.70-010800.9$ & 2007 Aug 6 & 2002 Apr 16 & 2008 Mar 1\\
\enddata
\end{deluxetable}

\clearpage
\begin{landscape}
\begin{deluxetable}{lrrrrrrrrrrrrrr}
\tabletypesize{\scriptsize}
\tablecolumns{15}
\tablewidth{0pc}
\tablecaption{UV and X-Ray Radio-Quiet Source Properties\label{uVXRayRQPropertiesTab}}
\tablehead{\colhead{SDSS DR5 Name} & \colhead{Redshift} & \colhead{ObsId}\tablenotemark{a} & \colhead{Expo} & \colhead{$AI$} & \colhead{$BI_0$} & \colhead{$v_{min}$} & \colhead{$v_{max}$} & \colhead{$\log(L_{\nu}(2500~\mathring{A}))$\tablenotemark{b}} & \colhead{$SN_{1700}$} & \colhead{$\log(L_{\nu}(2~{\rm keV}))$\tablenotemark{b,c}} & \colhead{HR\tablenotemark{d}} & \colhead{$\Gamma$} & \colhead{$\alpha_{OX}$} & \colhead{$\Delta\alpha_{OX}$} \\ \colhead{(J2000)} & \colhead{$z$} & \colhead{} & \colhead{(ks)} & \colhead{(km~s$^{-1}$)} & \colhead{(km~s$^{-1}$)} & \colhead{(km~s$^{-1}$)} & \colhead{(km~s$^{-1}$)} & \colhead{} & \colhead{} & \colhead{} & \colhead{} & \colhead{} & \colhead{} & \colhead{} }
\startdata
J$000654.10-001533.4$ & $1.725$ & 4096 &    4.5 & $  0.0$ & $  0.0$ &  &  & $31.209$ & $16.0$ & $26.960 \pm 0.091$ & $-0.55 \pm 0.23$ & $1.99 \pm 0.30$ & $-1.631 \pm 0.035$ & $0.034 \pm 0.035$\\
J$000659.28-001740.8$ & $2.016$ & 4096 &    4.5 & $  0.0$ & $  0.0$ &  &  & $30.790$ & $7.3$ & $26.851 \pm 0.106$ & $-0.60 \pm 0.32$ & $1.70 \pm 0.29$ & $-1.512 \pm 0.042$ & $0.094 \pm 0.042$\\
J$000856.79+155045.7$ & $1.691$ & 7127 &    4.0 & $  0.0$ & $  0.0$ &  &  & $30.954$ & $11.2$ & $26.903 \pm 0.108$ & $-0.70 \pm 0.30$ & $1.58 \pm 0.31$ & $-1.555 \pm 0.042$ & $0.074 \pm 0.042$\\
J$001247.12+001239.5$ & $2.153$ & 4829 &    6.7 & $  0.0$ & $  0.0$ &  &  & $31.047$ & $8.8$ & $26.691 \pm 0.194$ & $-0.57 \pm 0.45$ & $1.20 \pm 0.41$ & $-1.672 \pm 0.062$ & $-0.030 \pm 0.062$\\
J$001306.15+000431.9$ & $2.165$ & 4829 &    6.7 & $445.9$ & $  0.0$ & $-12320$ & $-13698$ & $31.229$ & $11.6$ & $26.526 \pm 0.165$ & $-0.66 \pm 0.42$ & $1.26 \pm 0.38$ & $-1.805 \pm 0.063$ & $-0.138 \pm 0.063$\\
J$002028.96+153435.8$ & $1.764$ & 1595 &   19.9 & $  0.0$ & $  0.0$ &  &  & $31.058$ & $12.2$ & $26.928 \pm 0.047$ & $-0.51 \pm 0.12$ & $1.61 \pm 0.16$ & $-1.585 \pm 0.018$ & $0.058 \pm 0.018$\\
J$002155.52+001434.3$ & $1.831$ & 0407030101 &   27.3 & $673.6$ & $  0.0$ & $-6620$ & $-14548$ & $30.544$ & $4.5$ & $26.143 \pm 0.103$ & $-0.63 \pm 0.18$ & $0.54 \pm 0.28$ & $-1.689 \pm 0.037$ & $-0.118 \pm 0.037$\\
J$002331.21-011045.6$ & $2.161$ & 4079 &    1.9 & $1974.0$ & $ 48.8$ & $-1041$ & $-20269$ & $30.682$ & $1.5$ & $<26.842$ &  &  & $<-1.474$ & $<0.117$\\
\enddata
\tablenotetext{a}{Four-digit ObsIds indicate {\it Chandra} exposures; ten-digit ObsIds indicate {\it XMM-Newton} exposures.}
\tablenotetext{b}{Monochromatic luminosity is in erg~s$^{-1}$~Hz$^{-1}$.}
\tablenotetext{c}{Errors are estimated using the method in \S\ref{xRayDataRedSec}.}
\tablenotetext{d}{Errors on hardness ratios are computed formally.}
\end{deluxetable}
\clearpage
\end{landscape}

\begin{deluxetable}{lrrrrrrrr}
\tabletypesize{\scriptsize}
\tablecolumns{9}
\tablewidth{0pc}
\tablecaption{Relative X-Ray Brightness Properties\label{sampleLumPropertiesTab}}
\tablehead{\colhead{} & \colhead{Number of} & \colhead{Number} & \colhead{$\log(L_{2500})$} & \colhead{$\log(L_{2500})$} & \colhead{$\Delta\alpha_{OX}$} & \colhead{$\Delta\alpha_{OX}$} & \colhead{$\alpha_{OX}$} & \colhead{$\alpha_{OX}$} \\ \colhead{} & \colhead{Sources} & \colhead{Detected} & \colhead{Median} & \colhead{Mean} & \colhead{Median} & \colhead{Mean} & \colhead{Median} & \colhead{Mean}}
\startdata
Non-BAL & 166 & 156 & $30.93$ & $30.98 \pm 0.03$ &  $0.01$ & $0.00 \pm 0.01$ & $-1.62$ & $-1.63 \pm 0.01$\\
Mini-BAL & 48 & 40 & $30.87$ & $30.89 \pm 0.06$ &  $-0.04$ & $-0.05 \pm 0.03$ & $-1.65$ & $-1.68 \pm 0.03$\\
BAL & 42 & 31 & $31.04$ & $31.09 \pm 0.06$ &  $-0.23$ & $-0.22 \pm 0.04$ & $-1.89$ & $-1.88 \pm 0.04$\\

\enddata
\end{deluxetable}

\begin{deluxetable}{lrr}
\tabletypesize{\scriptsize}
\tablecolumns{3}
\tablewidth{0pc}
\tablecaption{Hard X-Ray Power Law Photon Indices\tablenotemark{a}\label{sampleGammaPropertiesTab}}
\tablehead{\colhead{} & \colhead{$\Gamma$} & \colhead{$\Gamma$} \\ \colhead{} & \colhead{Median} & \colhead{Mean} }
\startdata
Non-BAL & $1.70$ & $1.59 \pm 0.08$\\
Mini-BAL & $1.55$ & $1.46 \pm 0.18$\\
BAL & $0.63$ & $0.87 \pm 0.16$\\

\enddata
\tablenotetext{a}{Sample properties calculated using only \mbox{X-ray} detected sources observed with {\it Chandra}.  The number of undetected sources is small (see Table~\ref{sampleLumPropertiesTab}), so we expect that the full sample properties should not differ greatly.}
\end{deluxetable}

\clearpage
\begin{landscape}
\begin{deluxetable}{lrrrlrrrrr}
\tabletypesize{\scriptsize}
\tablecolumns{10}
\tablewidth{0pc}
\tablecaption{{\it XMM-Newton} Observations of Selected Mini-BAL QSOs\label{makeXMMObsTableTab}}
\tablehead{\colhead{SDSS DR5 Name} & \colhead{Redshift} & \colhead{$pn$ Expo} & \colhead{$\log(R^*)$} & \colhead{Absorbed} & \colhead{T06 AI\tablenotemark{a}} & \colhead{$\log(L_{\nu}(2500~\mathring{A}))$\tablenotemark{b}} & \colhead{$\log(L_{\nu}(2~{\rm keV}))$\tablenotemark{b,c}} & \colhead{$\alpha_{OX}$} & \colhead{$\Delta\alpha_{OX}$} \\ \colhead{(J2000)} & \colhead{$z$} & \colhead{(ks)} & \colhead{} & \colhead{Ion} & \colhead{(km s$^{-1}$)} & \colhead{} & \colhead{} & \colhead{} & \colhead{} }
\startdata
J$111914.32+600457.2$ & 2.646 & 8.53 & 2.51 & \ion{C}{4} & 270.0 & 31.967 & $27.598 \pm 0.011$ & $-1.676 \pm 0.004$ & 0.094\\
J$145722.70-010800.9$ & 2.231 & 1.48 & $<$0.54 & \ion{C}{4} & 315.0 & 31.493 & $26.972 \pm 0.217$ & $-1.735 \pm 0.083$ & $-0.031$\\
\enddata
\tablenotetext{a}{Absorption index $AI$ reported by \citet{thrrsvkafbkn06}.}
\tablenotetext{b}{Monochromatic luminosity is in erg~s$^{-1}$~Hz$^{-1}$.}
\tablenotetext{c}{Errors are calculated using the method described in \S\ref{xRayDataRedSec}.}
\end{deluxetable}
\clearpage
\end{landscape}

% -------------- BEGIN FIGURES -----------------

\clearpage

\end{document}